\documentclass[apj]{emulateapj}

\usepackage{epsfig}
\usepackage{graphicx}
\usepackage{amsmath}
\usepackage{natbib}
\usepackage{aas_macros}
 \usepackage{hyperref}

\def\HII{{\ion{H}{2}}}

\def\4959_5007{[\ion{O}{3}]~$\lambda \lambda$4959,5007}
\def\OIII49595007{[\ion{O}{3}]~$\lambda \lambda 4959,5007$}
\def\ratioR23{([\ion{O}{2}]~$\lambda \lambda$3727,9 + [\ion{O}{3}]~$\lambda\lambda$4959,5007)/H$\beta$}
\def\R23{${\rm R}_{23}$}
\def\dS23{${\rm S}_{23}$}
\def\lOI{[\ion{O}{1}]~$\lambda$6300}
\def\lOII{[\ion{O}{2}]~$\lambda$3737,9}
\def\lOIII{[\ion{O}{3}]~$\lambda$5007}
\def\lNII{[{\ion{N}{2}}]~$\lambda$6584}

\def\ratioS23{([\ion{S}{2}]~$\lambda \lambda$6717,31 +[\ion{S}{3}]~$\lambda\lambda$9069,9532)/H$\beta$}

\def\O4363{[{\ion{O}{3}}]~$\lambda$4363}
\def\OIII{[{\ion{O}{3}}]}
\def\OIIIHb{[\ion{O}{3}]/H$\beta$}

\newcommand{\lzifu} {{\scshape lzifu}}
\newcommand{\ppxf} {{\scshape ppxf}}

\shorttitle{Multiple shocks in NGC~1052}
\shortauthors{Dopita et al.}

\begin{document}

\title{Probing the Physics of Narrow Line Regions in Active Galaxies III:  Accretion and Cocoon Shocks in the LINER NGC~1052}

\author{Michael A. Dopita\altaffilmark{1,2}, I-Ting Ho\altaffilmark{3}, Linda L. Dressel\altaffilmark{4}, Ralph Sutherland\altaffilmark{1},  Lisa Kewley\altaffilmark{1,4}, Rebecca Davies\altaffilmark{1}, Elise Hampton\altaffilmark{1}, Prajval Shastri\altaffilmark{6}, Preeti Kharb\altaffilmark{5}, Jessy Jose\altaffilmark{5},  Harish Bhatt \altaffilmark{5}, S. Ramya  \altaffilmark{5}, Julia Scharw\"achter\altaffilmark{6}, Chichuan Jin\altaffilmark{7}, Julie Banfield\altaffilmark{8},  Ingyin Zaw\altaffilmark{9}, Bethan James\altaffilmark{10}, St\'ephanie Juneau\altaffilmark{11},  \& Shweta Srivastava\altaffilmark{12}}
\email{Michael.Dopita@anu.edu.au}

\altaffiltext{1}{Research School of Astronomy and Astrophysics, Australian National University, Canberra, ACT 2611, Australia.}
\altaffiltext{2}{Astronomy Department, King Abdulaziz University, P.O. Box 80203, Jeddah, Saudi Arabia.}
\altaffiltext{3}{Institute for Astronomy, University of Hawaii, 2680 Woodlawn Drive, Honolulu, HI 96822, USA}
\altaffiltext{4}{Space Telescope Science Institute, 3700 San Martin Drive, Baltimore, MD 21218, USA}
\altaffiltext{5}{Indian Institute of Astrophysics, Koramangala 2B Block, Madiwala, Bangalore, 560034 Karnataka, India}
\altaffiltext{6}{LERMA, Observatoire de Paris, CNRS, UMR 8112, 61 Avenue de l'Observatoire, 75014, Paris, France}
\altaffiltext{7}{Qian Xuesen Laboratory for Space Technology, Beijing, China }
\altaffiltext{8}{CSIRO Astronomy \& Space Science, P.O. Box 76, Epping NSW, 1710 Australia }
\altaffiltext{9}{New York University (Abu Dhabi) , 70 Washington Sq. S, New York, NY 10012, USA }
\altaffiltext{10}{Institute of Astronomy, Cambridge University, Madingley Road, Cambridge CB3 0HA, UK }
\altaffiltext{11}{CEA-Saclay, DSM/IRFU/SAp, 91191 Gif-sur-Yvette, France}
\altaffiltext{12}{Astronomy and Astrophysics Division, Physical Research Laboratory, Ahmedabad 380009, India}

\begin{abstract}
We present Wide Field Spectrograph (WiFeS) integral field spectroscopy and HST FOS spectroscopy for the LINER galaxy NGC~1052. We infer the presence of a turbulent accretion flow forming a small-scale accretion disk. We find a large-scale outflow and ionisation cone along the minor axis of the galaxy. Part of this outflow region is photoionised by the AGN, and shares properties with the ENLR of Seyfert galaxies, but the inner ($R \lesssim 1.0$~arcsec) accretion disk and the region around the radio jet appear shock excited. The emission line properties can be modelled by a ``double shock" model in which the accretion flow first passes through an accretion shock in the presence of a hard X-ray radiation, and the accretion disk is  then processed through a cocoon shock driven by the overpressure of the radio jets. This model explains the observation of two distinct densities ($\sim10^4$ and $\sim10^6$~cm$^{-3}$), and provides a good fit to the observed emission line spectrum.  We derive estimates for the velocities of the two shock components and their mixing fractions, the black hole mass, the accretion rate needed to sustain the LINER emission and derive an estimate for the jet power. Our emission line model is remarkably robust against variation of input parameters, and so offers a generic explanation for the excitation of LINER galaxies, including those of spiral type such as NGC~3031 (M81).
\end{abstract}


\section{Introduction}\label{sec:intro}
Since the low-ionisation nuclear emission line region (LINER) class of galaxies was originally formalised by \citet{Heckman80}, the nature of their excitation has remained a cause of speculation. \citet{Heckman80} distinguished these galaxies from Seyfert galaxies and from galaxies with \HII-region-like emission lines with the following criteria: \lOII $ \gtrsim$ \lOIII, and  \lOI $ \gtrsim$  \lOIII. The surveys by \citet{Oconnell78} and \citet{Heckman80} confirmed what had previously been suspected from a few coincidences: a strong correlation between bright LINER emission and powerful compact nuclear radio sources. In most of these bright LINERs, virtually all of the radio emission comes from flat-spectrum compact self-absorbed synchrotron sources \citep{Condon78}. Strong LINER emission is also sometimes seen in powerful radio galaxies with extended steep-spectrum synchrotron emission, such as M87 \citep{Ford94,Harms94}.
 
Low level LINER emission has now been detected in a large fraction of elliptical galaxies \citep{Phillips86, Veron-Cetty86, Ho96,Ho97}. This emission may have a variety of origins, nonetheless the bright LINERs associated with powerful radio sources appear to be true AGN. Both their radio emission, and the presence of broad H$\alpha$ emission \citep{Ho97b,Younes12} suggests a link to radio-powerful Seyfert galaxies and quasars. A few LINERs have been observed to have strong, compact and spectrally-hard X-ray emission implying that the source is an AGN (Serlemitsos et al 1996). Indeed, \citet{Kewley06} found that the nuclear LINERs in red emission-line SDSS galaxies lie along the same host-galaxy Eddington ratio relations as Seyferts, which also implies an AGN origin for these LINERS. Energy considerations suggest that the power source is the release of gravitational energy by matter dropping into a supermassive black hole. Kinematic measurements, beginning with the LINER M87 \citep{Ford94,Harms94}, show that the velocity widths can be understood as primarily due to orbital motion around a massive black hole. However, even the ``narrow line'' components in LINERs have FWHM of hundreds of km~s$^{-1}$ when observed through HST apertures. This is broader than expected from the variation of rotational velocity across a small aperture placed off-nucleus, and implies the presence of turbulence, non-rotational flows, or high velocity shocks.

The optical emission line ratios of LINERs were originally  modelled as being due to shock-heated gas \citep{Baldwin81}.  It was soon discovered, however, that very similar line ratios could also be produced by gas that was photoionized by either a power-law or a thermal Bremstrahhlung continuum with a low ionization parameter; i.e., a low ratio of incident ionizing photons per ion  \citep{Ferland83}. To add to the possible physical models, \citet{Terlevich85} pointed out that a similar hard UV continuum could be produced by very hot luminous Wolf-Rayet stars in a nuclear starburst. 

Strictly speaking, the term LINER should be confined to describing the nuclear region of a galaxy - according to the original definition by \citet{Heckman80}. However, many host galaxies show extended emission on kiloparsec scales which maintains a LINER-like spectrum. Many of the objects with LINER-like spectra found in surveys are of this nature, and it may well be that this extended emission is excited in a different manner from the nuclear emission. Such regions were distinguished from the nuclear emission line regions as ``extended LINERs" and were analysed by \citet{Yan12}. They found that the extended emission is inconsistent with ionisation from a central object, and that rather, extended sources such as post-AGB stars are required to provide the ionisation; an idea originally proposed by \citet{Binette94}, and recently developed in the light of CALIFA  integral field spectroscopic data by \citet{Singh13}. These authors conclude that extended LINER emission of itself cannot be use to infer the presence of an AGN. 

From the velocity dispersions and the small width differences between different lines,  \citet{Yan12} concluded that the line width in extended LINERs is dominated by bulk motion and turbulence, and that two ISM components are needed to produce line width differences between different ions. This is perhaps difficult to reconcile with ionisation by the post-AGB component alone. However, \citet{Farage10} showed that shocks in a multi-phase medium during a merger event can produce different line widths in different ions. Furthermore, in first-ranked Ellipticals \citep{Farage12}, the observed line ratios show (apparently) more shock dominance closer to the nucleus, again showing how difficult it is to interpret the extended LINER emission using a single mechanism.

Significant progress in LINER emission line diagnostics can be made by extending the flux predictions for power law, bremsstrahlung, and shock models into the UV. In the case of M87, \citet{Dopita97} have shown that several UV lines are much stronger in shock models than in the other types of model. Unfortunately, the UV spectral data is available for relatively few LINERs. Such data as do exist \citep{Maoz98} seem to show that the LINER class is not homogenous in its UV properties, with some objects exhibiting strong emission lines, and others having a UV spectrum that is simply consistent with an old stellar population.

\begin{figure}[htb!]
\begin{centering}
\includegraphics[scale=0.45]{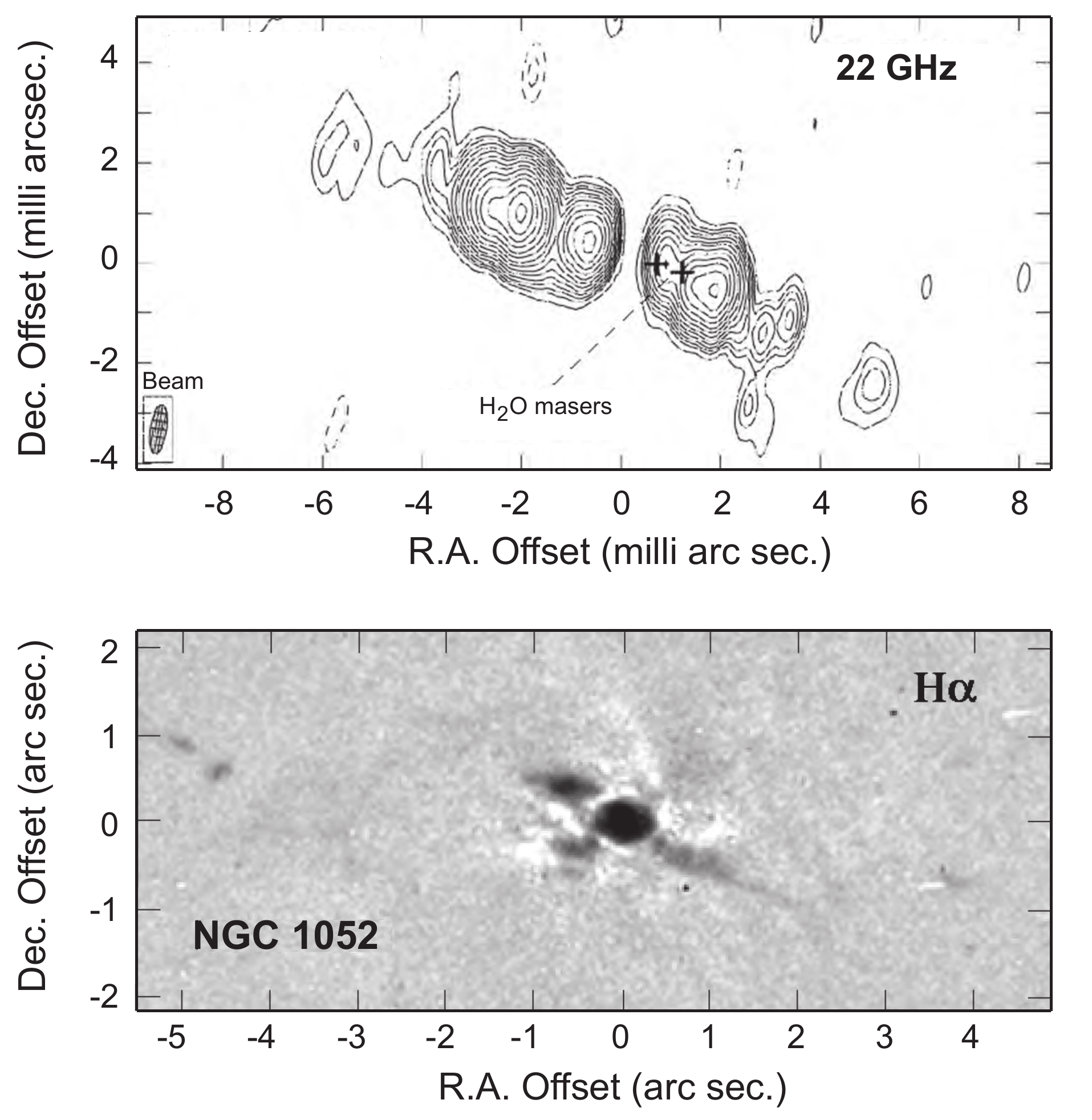}
\end{centering}
\caption{Upper panel, from \citet{Claussen98} shows the radio jet and the H2O masers in NGC 1052. One mas corresponds to a distance of about 18000 AU. The masers are probably associated with a dense circum-nuclear disk, and are formed by amplification of the background radio continuum emission. The lower panel shows the HST image of the (galaxy subtracted) H$\alpha$ jet - perfectly aligned with the inner radio structure, but about 500 times longer. This structure apparently represents the interaction of an outflow with the circum-nuclear ISM. The white regions are regions of high dust extinction and may delineate the shock-compressed cocoon of the radio jet. At larger distances $(\sim$ 4~arcsec) from the nucleus there are a number of bright knots, which are part of a filamentary complex best seen in fig. 2 of \citet{Pogge00}. Figure courtesy of Dr. Andrew Wilson, deceased.}\label{fig:NGC1052_jet}
\end{figure}

NGC 1052 is an E4 galaxy, is the prototypical LINER galaxy, and being nearby (z= 0.0045) can be studied in detail. It is a radio galaxy having a well-collimated non-thermal radio jet on both ~arcsec scales \citep{Wrobel84,Cooper07} and milli-arcsec scales \citep{Claussen98}. The radio jet shows sub-relativistic expansion with $\beta \sim 0.24-0.38$ \citep{Kadler04b,Lister13}.

However, it also has a dense nuclear accretion disk, demonstrated by the fact that it is one of only two elliptical galaxies containing very luminous water masers. The H$_2$O maser emission lines are relatively broad (90~km~s$^{-1}$) and smooth, unlike the narrow maser spikes seen in other active galaxies \citep{Braatz96}. VLBI observations by \citet{Claussen98} (see fig. \ref{fig:NGC1052_jet}) show that the masers lie along, rather than perpendicular to, the jet, and only on the west south-west side of the nucleus. The scale of the maser emission is about 400 $\mu$as ($\sim$7000 AU). \citet{Claussen98} argue that although the jet is energetically capable of powering the observed maser emission by driving slow, non-dissociative shocks into circum-nuclear, dense molecular clouds, it is more likely, given their location and velocity structure, that the masers represent the direct amplification of the radio continuum emission of the jet by the foreground molecular clouds in an inclined nuclear accretion disk.

At X-ray wavelengths, the nucleus appears as a hard, heavily obscured ($N_{\rm H} \approx 10^{23}$~cm$^{-2}$), flat-spectrum ($\alpha \sim 0.1$) point source with an (unabsorbed) luminosity of $5\times 10^{41}$erg~s$^{-1}$ \citep{Guainazzi99, Kadler04b}. With \emph{CHANDRA} observations,  a more extended region of thermal X-ray emission ($kT \sim 0.5$~keV) is seen occupying a double wedge-shaped region extending $\approx 10$~arcsec. east and west of the nucleus \citep{Kadler04b}. Although its extent is similar to the radio lobes observed by  \citet{Wrobel84}, in detail, \citet{Kadler04b} finds an anti-correlation between the X-ray and radio surface brightness, suggesting that the X-ray gas may be associated with the working surface of the radio lobes.

The optical emission line gas (see fig. \ref{fig:NGC1052_jet}) shows a jet component extended on~arcsec scales, perfectly aligned with the innermost radio jet, possibly ionised by UV synchrotron emission beamed from the inner jet. Elsewhere there are H$\alpha$ emitting blobs and filaments filling the zone where the thermal X-rays are observed. However, the most prominent feature as well as a bright nuclear region 100-200~pc in diameter \citep{Pogge00}. This has been studied in detail by \citet{Sugai05}, and probably represents a resolved accretion disk. The line flux coming from the much larger extended region is similar in total luminosity to that of this inner disk.

In some sense, NGC~1052 could be classified as a ``hidden broad-line LINER"  following the detection of a prominent broad H$\alpha$ component with FWHM $\sim 5000$~km~s$^{-1}$ in polarised light \citep{Barth99}. Broad wings are also seen in un-polarised light. This data establishes the existence of a dense inner accretion disk with a hot scattering region around it.

The interpretation of the emission line spectrum of NGC~1052 (like other LINERs) remains controversial. The idea that the NGC~1052 emission is shock excited was originally suggested by \citet{Koski76}, and a detailed radiative shock model was presented by \citep{Fosbury78}. However, this model fell out of favour following the apparent success of photoionisation models. However, from infrared data, \citet{Sugai00} found evidence that shock excitation might well provide a substantial fraction of the total emission in NGC~1052. In contradiction to this explanation, an examination of the HST data by \citet{Gabel00} (re-analysed here) suggested that the the emission-line fluxes can be simulated with a simple photoionization model using a central power-law continuum source with a spectral index of $\alpha = -1.2$. Two components with radically different densities and ionisation parameters were required in this model, leaving the physical origin of these two components something of a mystery.

A vital clue to the resolution of this issue was provided by the sub-~arcsecond resolution integral field observations of \citet{Sugai05} who used the Kyoto Tridimensional Spectrograph II mounted on the Subaru Telescope. They found evidence for a high-velocity bipolar outflow, a low-velocity rotating disk and a spatially unresolved broad line region visible in H$\beta$. The high velocity, outflowing [\ion{O}{3}] emission ridges closely correlate with the radio structures mapped with MERLIN at 1.4GHz by \citet{Kadler04a}. Whilst  \citet{Sugai05} interpret these observations in terms of a directed outflow, we argue here that it is more likely that this emission is caused by a cocoon shock as advocated by \citet{Bicknell97} in the context of gigahertz-peaked spectrum (GPS) and compact steep-spectrum (CSO) radio sources and by \citet{Bicknell98} in the context of Seyfert galaxies. 

In this paper we argue that, in NGC~1052, such cocoon shocks are propagating into the (already shocked) accretion disk, giving rise to the dense emission component required by the HST observations. In addition, we present integral field spectroscopy of the central $25\times38$\~arcsec\  region of NGC~1052 which establishes the existence of a higher excitation ``ionisation cone'' and a large-scale bipolar mass outflow presumably energised by the radio jet. In addition we will show the existence of a turbulent accretion flow along the major axis of the galaxy, and in the plane of rotation of the stars. In order to understand the excitation of the inner, bright  H$\alpha$ region, we re-examine the HST data of \citet{Gabel00}, in the light of the observations of  \citet{Sugai05} and construct a new accretion shock plus cocoon shock model which naturally explains the existence of the two density components referred to above.

The paper is organised as follows. In Section \ref{Observations} we present the data, and the extraction of the reduced quantities. In Section \ref{sec:shocks} we introduce the physical basis of our multiple shock model to explain the spectrum of NGC~1052. In Section \ref{sec:model} we present our accretion + cocoon shock model in the presence of a hard X-ray spectrum which is very successful in reproducing the observed LINER spectrum of NGC~1052 over a wide range of assumed shock velocities in either component. In Section \ref{discussion} we use the observations to estimate the black hole mass, the mechanical energy luminosity of the radio jet, and the mass accretion rate into the accretion disk. We assume the redshift-independent distance to NGC~1052 given by the NASA Extragalactic Database (NED); 19.9~Mpc which implies a spatial scale of 97 pc ~arcsec$^{-1}$.

\section{Observations and Results}\label{Observations}
\subsection{The WiFeS integral field observations}
NGC~1052 was observed on 2-3 November 2013 using the Wide Field Spectrograph \citep[WiFeS;] []{Dopita07,Dopita10} at the ANU 2.3m telescope at Siding Spring Observatory. WiFeS is an optical integral field spectrograph providing a field of view of 25~arcsec $\times$ 38~arcsec via a total of twenty five  38~arcsec $\times$ 1~arcsec slitlets. The instrument is a double-beam spectrograph, with arms for both the blue and for the red part of the spectrum. Thus, a  wide wavelength range in the optical is covered simultaneously, and with adequate spectral overlap between the red and blue spectra. 

\begin{figure}[htb!]
\begin{centering}
\includegraphics[scale=0.55]{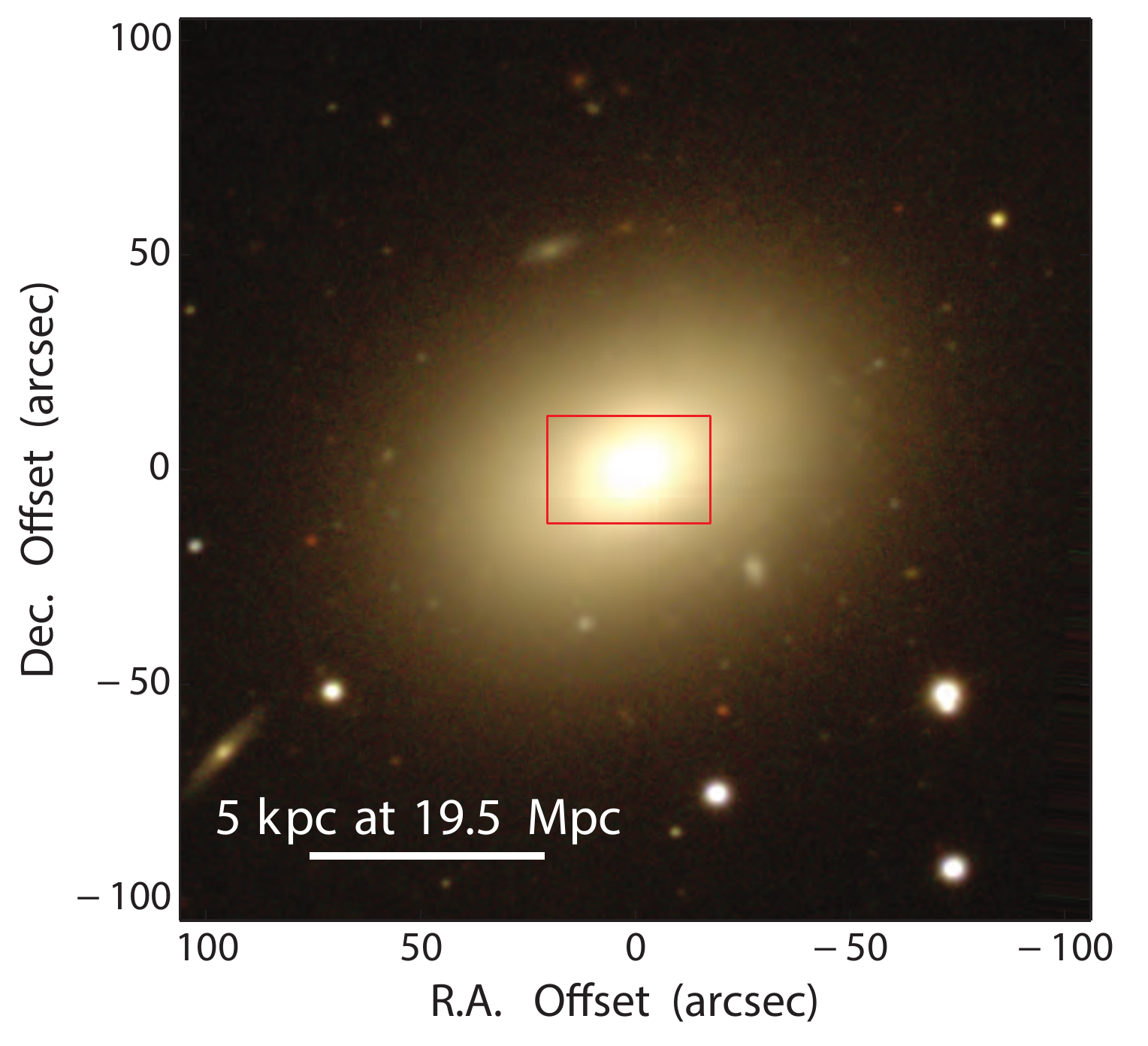}
\end{centering}
\caption{The SDSS {\it g,r,i} image of NGC~1052 with the WiFeS pointing position superimposed. We also show the scale at the assumed distance of 19.9~Mpc. }\label{fig:point}
\end{figure}

The observations for NGC~1052 consisted of $3\times1000$s exposures at RA = 02:41:04.9 Dec= -08:15:21.1 (J2000) and $3\times1000$s exposures on the nearby sky at the position RA = 02:41:10.5 Dec= -08:11:41.4 (J2000). All observations were obtained using the $R_S = 3000$ grating in the blue ($B3000$ grating) and the $R_S = 7000$ grating in the red arm ($R7000$ grating). The position of the WiFeS data cube is shown on an image of NGC~1052 in Figure \ref{fig:point} with the physical scale at the assumed distance of 19.5~kpc. Absolute photometric calibration was made using the STIS spectrophotometric standard stars HD~009051 and HD~031128 \footnote{Available at : \newline {\tt www.mso.anu.edu.au/~bessell/FTP/Bohlin2013/GO12813.html}}. In addition two B-type telluric standards were observed, HIP~108975 and HIP~18926. During the observation, the seeing averaged 1.3-1.5~arcsec, reasonably well matched to the 1.0~arcsec pixels of the spectrograph. \bigskip

\subsubsection{WiFeS Data reduction}

The data were reduced using the {\tt PyWiFeS} pipeline written for the instrument \citep{Childress14}. In brief, this produces a data cube which has been wavelength calibrated, sensitivity corrected (including telluric corrections), photometrically calibrated, and from which the cosmic ray events have been removed.

For the resultant data cube, we have used the IFS toolkit \lzifu\ (Ho et al. 2014 in prep.) to extract gas and stellar kinematics from the WiFeS data. \lzifu\ uses the the penalized pixel-fitting routine \citep[\ppxf\;][]{Cappellari04} to perform simple stellar population (SSP) synthesis fitting to model the continuum, and fits the emission lines as Gaussians using the Levenberg-Marquardt least-squares technique \citep{Markwardt09}. For generating stellar velocity and velocity dispersion maps we employ the theoretical SSP libraries from \citet{Gonzalez05} assuming the Padova isochrones. For the gas velocity and velocity dispersion maps we simultaneously fit within each spaxel one-component Gaussians to 11 strong optical emission lines and we constrain all the lines to have the same velocity and velocity dispersion. The 11 lines of the fit are [\ion{O}{2}]$\lambda \lambda 3726,29$, H$\beta$,  [\ion{O}{3}]$\lambda \lambda 4959,5007$,  [\ion{O}{1}]$\lambda 6300$, [\ion{N}{2}]$\lambda \lambda 6548,83$, H$\alpha$, and [\ion{S}{2}]$\lambda \lambda 6716,31$. We fix  the ratios [\ion{O}{3}]$ \lambda \lambda 4959/5007$ and [\ion{N}{2}]$ \lambda \lambda 6548/6584$ to their theoretical values given by quantum mechanics \citep{ADU}. 

\begin{figure}[htb!]
\begin{centering}
\includegraphics[scale=0.4]{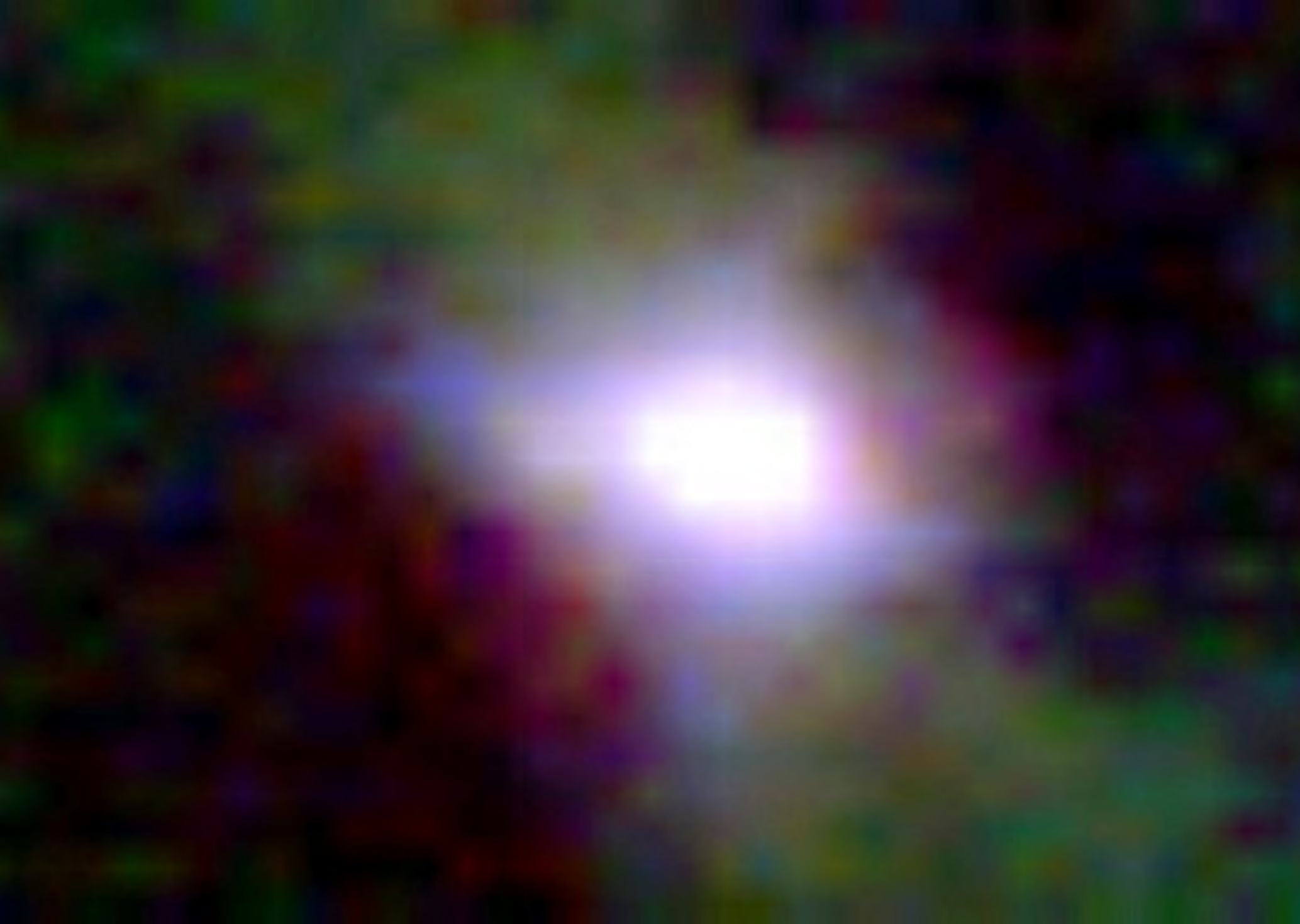}
\end{centering}
\caption{The $38\times25$~arcsec WiFeS image of NGC~1052 in [\ion{O}{3}] $\lambda 5007$ (blue), H$\alpha$ (green) and [\ion{N}{2}]  $\lambda 6584$ (red). The stretch has been chosen so as to enhance the fainter features, and the image has been smoothed with a rectangular box-car function so as to reduce pixelation. Note the E-W extension of the [\ion{O}{3}]  - bright central region along the direction of the arc sec. scale radio jets \citep{Wrobel84}, and the prominent X-shaped central ionisation cones in  [\ion{O}{3}]. H$\alpha$ is enhanced in a two-sided funnel in the direction of the minor axis of the galaxy (along the rotation axis of the stars). [\ion{N}{2}] is enhanced along the major axis in the region of extended thermal X-ray emission.}\label{fig:WiFeSimage}
\end{figure}

\subsubsection{Results from the Data Cube}
The WiFeS data cube furnishes a number of novel and interesting results. The emission line maps reveal that the degree of excitation is in no sense uniform. Indeed, in a restricted X - shaped zone extending nearly 10~arcsec from the nucleus, we find [\ion{O}{3}] lines to be strongly enhanced, approaching the values seen in Seyfert narrow line regions; see figure \ref{fig:WiFeSimage}. We regard this as probably delineating an ionisation cone photo ionised by the central engine. In the nuclear region an enhanced region of [\ion{O}{3}]  emission is seen to be associated with the arc sec. scale radio jet \citep{Wrobel84}, while H$\alpha$ is enhanced along the axis of rotation of the stars (the minor axis).  The [\ion{N}{2}] emission is relatively strongest along the major axis of the galaxy, within the X - shaped [\ion{O}{3}] ionisation cone in which the \OIIIHb\ ratio rises to 3-5. The spatial extent and shape of this region is very similar to the region within which the thermal X-rays are enhanced (\emph{c.f} \citet{Kadler04b}).
\begin{figure*}[htb!]
\begin{centering}
\includegraphics[scale=0.65]{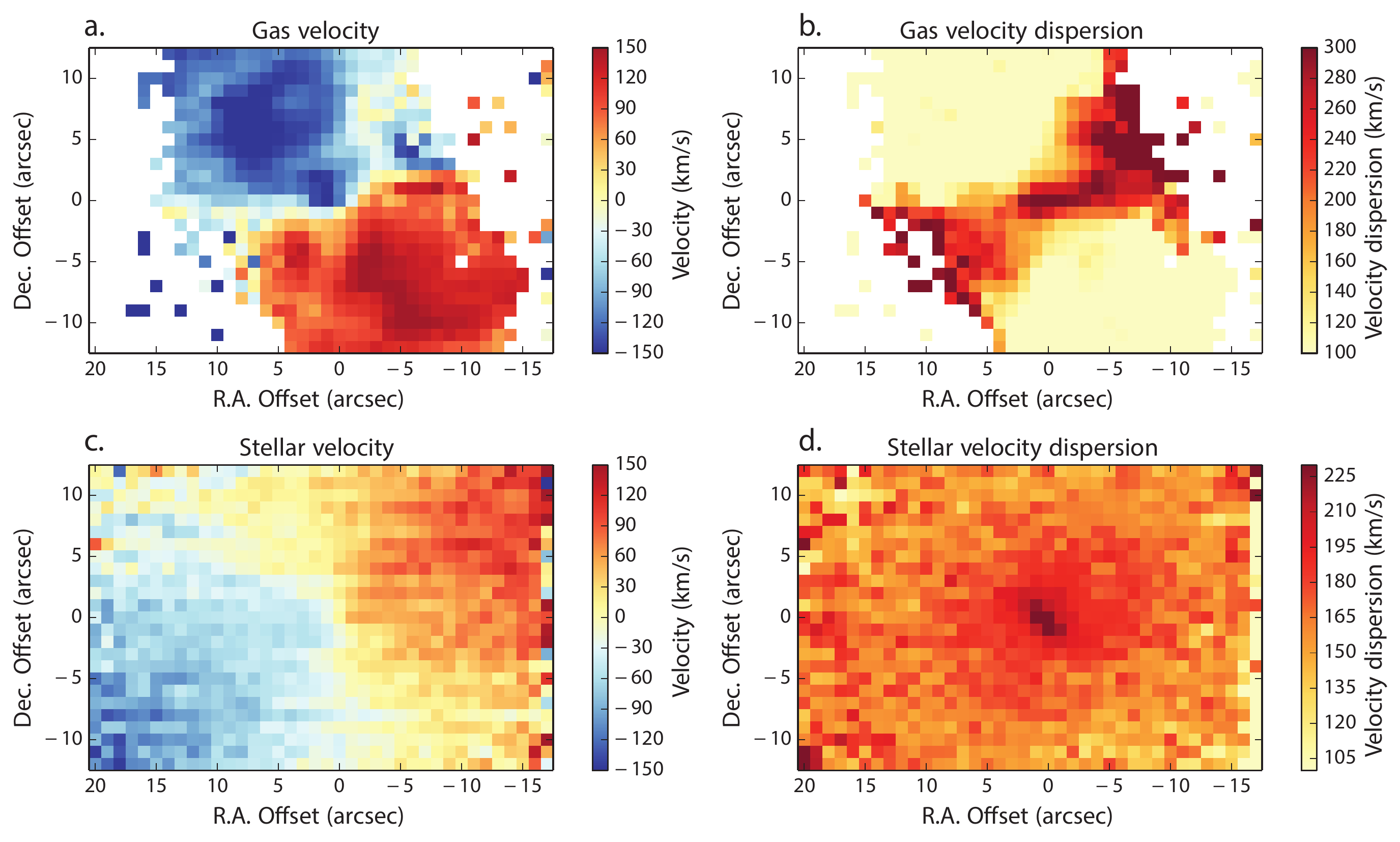}
\end{centering}
\caption{The dynamical structure if NGC~1052 in H$\alpha$ and in the stellar component. The H$\alpha$ velocity field (panel a.) shows two bubbles of gas expanding along the minor axis of the galaxy, separated by a region of very high velocity dispersion seen in panel b. The stellar dynamics are dynamically quiescent. The stars rotate slowly around the photometric minor axis (panel c.), and with a velocity dispersion of $\approx 200$km~s$^{-1}$, typical of the values observed in elliptical galaxies (panel d.).}\label{fig:WiFeSdynamics}
\end{figure*}

The dynamical picture -- shown in Figures \ref{fig:WiFeSdynamics} -- is equally interesting. The stars rotate smoothly around the photometric minor axis, and are characterised by a velocity  dispersion of $\approx 200$~km~s$^{-1}$, not unusual for an elliptical galaxy \citep{Gerhard01,Thomas07,Cappellari13}. There is a sharp cusp in the velocity dispersion close to the nucleus, presumably within the zone of influence of the central black hole. The dynamics of the gas is completely different to the stellar dynamics. The gas velocity component analysis reveals two bubbles of gas expanding along the minor axis of the galaxy, separated by a region of very high velocity dispersion. These bubbles are about 15\~arcsec\ in diameter ($\sim 1.5$~kpc), and are well-aligned with the H$\alpha$ emission ``funnel'' seen in Figure  \ref{fig:WiFeSimage}. The bubble on the NE side is approaching, so this is located on the near side of the galaxy.  This is consistent with the fact that the H$_2$O maser emission associated with amplification of jet radio continuum by the inner accretion disk is on the opposite side of the nucleus, also suggesting that the ENE jet is directed somewhat towards us \citep{Claussen98} (see fig. \ref{fig:NGC1052_jet}). The velocity of approach in the NE bubble, and the recession velocity in the SW bubble is greatest along the axis, but decreases strongly towards the boundary of the bubbles. In addition, the internal velocity field is irregular and quite unlike a rotation curve. Furthermore, both bubbles show an internal velocity dispersion of about 120~km~s$^{-1}$, much greater than the instrumental resolution in the red ($\sim 45$~km~s$^{-1}$. All these dynamical properties are consistent with the buoyant expansion of turbulent bubbles filled with hot or non-thermal plasma in the ISM of the galaxy.

The region between the two expanding H$\alpha$-emitting bubbles displays a very high velocity dispersion, up to 300~km~s$^{-1}$. In both extent and morphology it coincides with the region within which the thermal X-rays are enhanced \citep{Kadler04b} and the region in which the [\ion{N}{2}] emission is unusually strong with respect to H$\alpha$, referred to above.  The juxtaposition of high velocity dispersion, enhanced [\ion{N}{2}]/H$\alpha$ ratios, and the presence of thermal X-ray emission all suggest the presence of strong radiative shocks in a highly turbulent medium \emph{c.f} \citet{Ho14}. Its inner region also corresponds to the zone of enhanced obscuration seen in Figure \ref{fig:NGC1052_jet}.

\subsection{The FOS Observations}
The spectral data used in this study were obtained under project GO-6532 (investigators: Dopita \& Koratkar) using  the Faint Object Spectrograph (FOS) aboard the Hubble Space Telescope (HST /FOS) on 1997 January 13 (post-COSTAR). All observations were made using the 0.86~arc sec. pair square aperture and the wavelength coverage was 1200 - 6800\AA. The optical spectra (G400H and G570H grating) and the NUV (G270H grating) data were obtained at a resolution of $\lambda/\Delta\lambda \sim 1300$ (or an approximate resolution 2.1-4.4\AA). Because of the faintness of the lines, the UV spectrum was obtained with the  G160L grating giving a resolution of $\lambda/\Delta\lambda \sim 250$ (or an approximate resolution of 6.4\AA).  For NGC~1052, the aperture was well-matched in size to the central bright region visible in the H$\alpha$ image of Fig. \ref{fig:NGC1052_jet}, and so provides the integrated spectrum of the true nuclear LINER in this galaxy. 

\begin{figure}[htb!]
\begin{centering}
\includegraphics[scale=0.5]{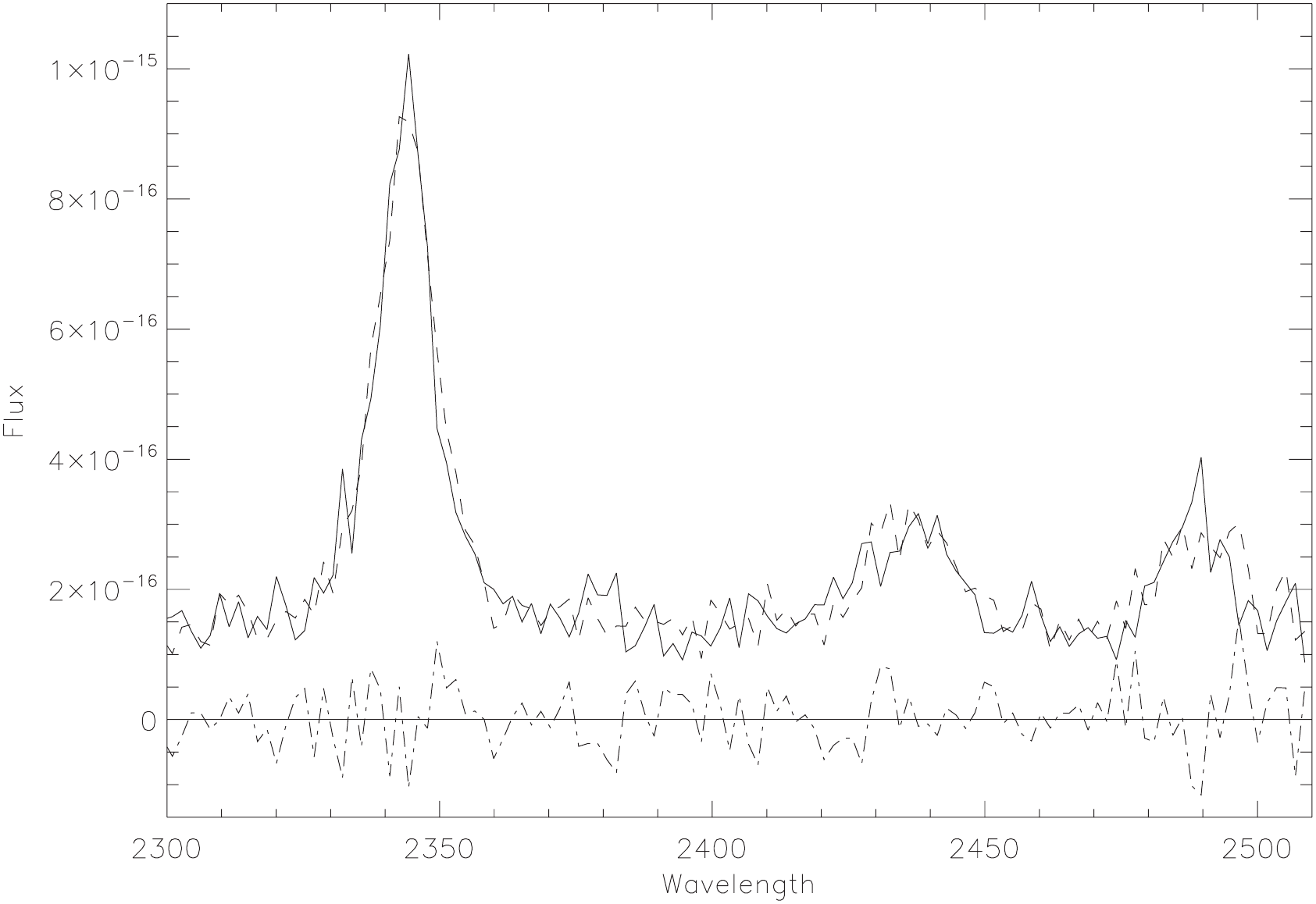}
\end{centering}
\caption{Archival spectra from two orbit-long G160L exposures (solid and dashed line) and their difference (dot-dashed line), compared for evidence of drift of the target. We may conclude that any such drift is negligible.}\label{fig:UVdifference}
\end{figure}

     The G160L data were taken in three exposures covering two and a half orbits.  These were followed by the G270H exposure in the next orbit and the G400H and G570H exposures in the final orbit.  Only one guide star was acquired at the outset, so the roll of the telescope was controlled by gyroscope, which generally results in a slow drift of the target across the detector over the course of several orbits.  Archival data for the two full-orbit G160L exposures were checked for systematic differences in the bright emission lines that would indicate substantial drift.  Figure \ref{fig:UVdifference} shows that such differences are small.
     
\subsubsection{Extracting the Emission line spectrum}
\begin{figure*}[htb!]
\begin{centering}
\includegraphics[scale=1.02]{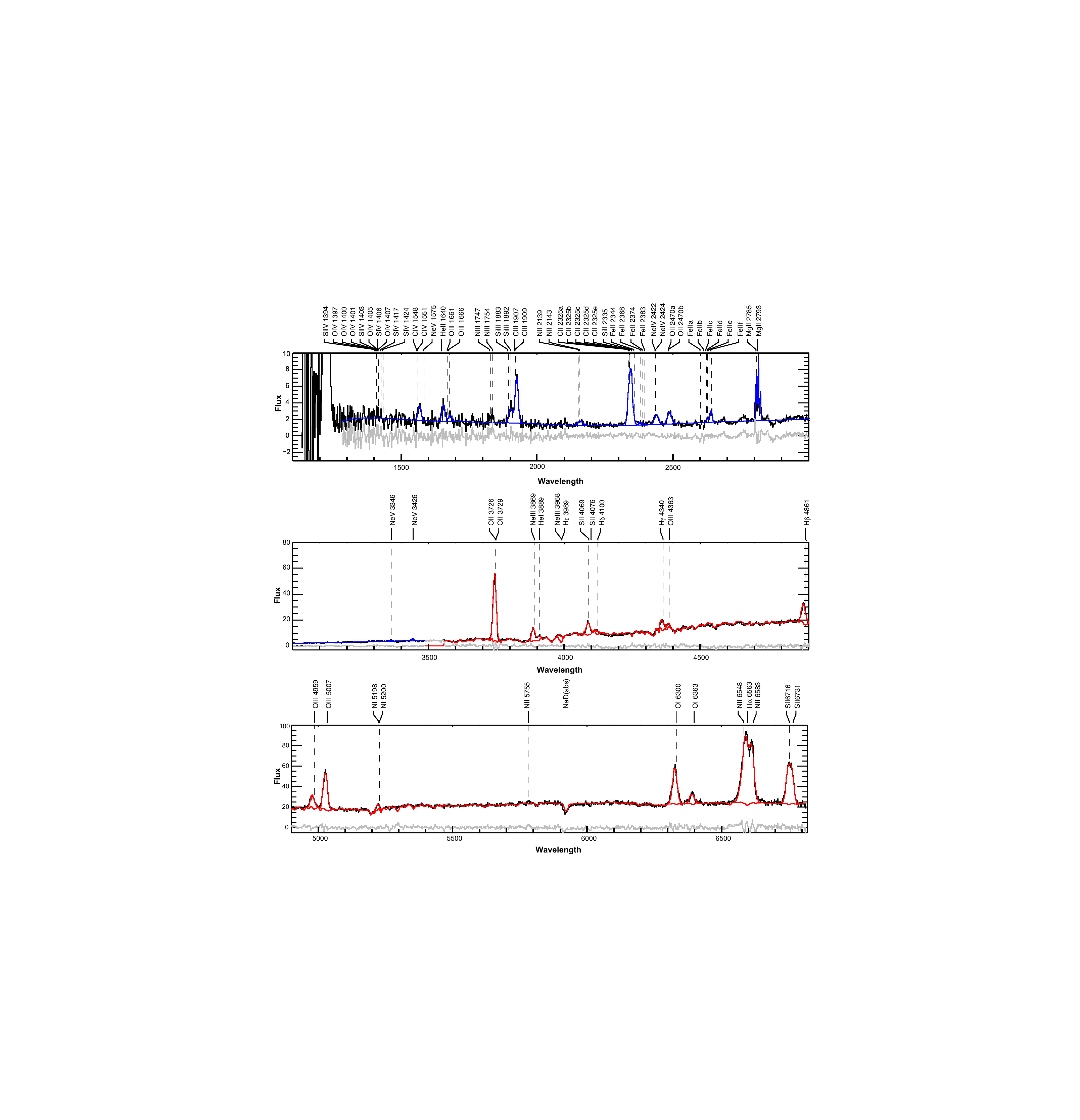}
\end{centering}
\caption{The HST FOS spectrum of NGC~1052 (solid black line), the fitted line and continuum spectrum (blue line for the UV or red line for optical wavelengths), and the residuals (gray line). For the continuum, a stellar template is used down to 3550\AA (thin red line), and a Chebyshev polynomial fit shortward of that (thin blue line).}\label{fig:NGC1052_HST}
\end{figure*}

Figure \ref{fig:NGC1052_HST} shows the FOS spectrum,  along with the continuum and line model fits used to extract the line fluxes. We use a modified version of \lzifu\ to fit the spectrum. For the continuum fitting, we adopt the MILES stellar libraries \citep{Vazdekis10} at wavelengths greater than $\lambda \approx3500$\AA. At wavelengths smaller than $\lambda \approx3500$\AA, we fit the continuum with linear combinations of 10 orders of Legendre polynomial. Channels around emission lines are masked out while fitting the continuum. For the emission line fits, we split the emission lines in 14 groups based on their wavelengths and then perform separate fits to each group. In each fit, we constrain all the lines to have the same velocity and velocity dispersion. All the emission lines are described by single component Gaussians, except for the H$\alpha$ line where a second Gaussian component is required to properly describe the line profile. Where possible, the flux ratios of the emission lines from the same multiplet of a given ion are constrained based on the relative probabilities of their atomic transitions.

\subsubsection{The FOS Emission Line spectrum}\label{FOSspectrum}
The de-reddened extracted emission line intensities are listed in Table 1. The line fluxes have been de-reddened by the theoretical function for a physical grain model derived by \citet{Fischera05} and which provides a very close fit to the Galactic attenuation law with $R=3.2$. In common with \citet{Gabel00} we find an extinction which is unusually high for the LINER population; $A_{\rm V} = 1.05$. This corresponds to E(B-V) = 0.33, somewhat lower than the value of E(B-V) = 0.44 estimated by  \citet{Gabel00}. The presence of such a large amount of circum-nuclear dust suggests that the nuclear region also contains a large amount of interstellar gas, which we will interpret as due to a dusty accretion flow onto the bright nuclear disk. Indeed, this dusty flow can be seen directly in Figure 4 of \citet{Kadler04a} as a complex of accretion streams within the central $\sim 2$~arcsec

The line fluxes of Table 1 cannot be fit by a simple one-zone photoionisation model. As realised by  \citet{Gabel00} the line ratios are inconsistent with such a model. While the [\ion{S}{2}]  $\lambda 6717,31$ line ratio and the presence of a strong  [\ion{O}{2}]  $\lambda 3726,9$ doublet suggest a density of order $300 \lesssim  n_e \lesssim 1000$~cm$^{-3}$, ratios such as [\ion{S}{2}]  $\lambda 4068,76/6717,31$,
 [\ion{N}{2}]  $\lambda 5755/6584$, [\ion{O}{2}]  $\lambda 2470/3726,9$ and [\ion{O}{3}]  $\lambda 4363/5007$ are all indicative of densities in the range  $10^6 \lesssim  n_e\lesssim 10^7$~cm$^{-3}$. \citet{Gabel00} model these two components as physically independent entities photoionised by the same power-law spectrum, and derive appreciably different ionisation parameters for the two components. Surprisingly the denser component has the higher ionisation parameter, so if the model is correct, it would have to be emitted much closer to the central engine. If so, why then does the dense component not show the broader line widths expected with the Keplerian motion around the central engine? In what follows we will present a physically motivated accretion model with photoionisation which appears to provide a natural answer to this question.

\section{Multiple shocks in NGC~1052}\label{sec:shocks}
\subsection{The Large Scale Flows}
The WiFeS observations of NGC~1052 clearly reveal the presence of a high velocity dispersion gas along the major axis of the galaxy, associated with enhanced  [\ion{N}{2}]  $\lambda 6584$ emission. In addition, this data also provides evidence for a jet-induced outflow with embedded ionisation cone along the minor axis of the galaxy. The high velocity dispersion along the major axis may be due to a variety of causes. First, it could be simply line of sight overlap between the two expanding bubbles. Second, it could be due to a physical interaction between the ambient medium and the radio jets. Third, it could be due to intrinsic velocity dispersion and shocks in a turbulent accretion flow. In all of these scenarios, shocks play a major role, and the presence of thermal X-ray emission co-extensive with the region of high velocity dispersion tends to support this picture. The temperature of this X-ray plasma ($kT = 0.4-0.5$~keV), if shock-heated, would require shocks with a velocity of $\sim 400$~km~s$^{-1}$. 

Figure \ref{fig:NGC1052_jet} reveals the presence of dusty (accretion?) streams as well as both clumpy and filamentary emission line gas in the region occupied by the thermal X-ray plasma. Figure \ref{fig:WiFeSimage} shows that the [\ion{N}{2}] 6584\AA\ emission is particularly strong in this region, which is bounded by the X-shaped region of high [\ion{O}{3}] 5007\AA\ emission which coincide with the inner boundary of the expanding large-scale bubbles of gas seen in Figure \ref{fig:WiFeSdynamics}.

We are thus led to a model which has both inflows and outflows on the large scale. An accretion flow is falling into a small few arc sec. diameter circum-nuclear region. The accretion flow is mostly confined within 45 degrees of the major axis of the galaxy. At the same time, a hot plasma consisting of both relativistic gas from the nucleus and a hot X-ray plasma with $kT \sim 0.4-0.5$~keV is rising buoyantly at velocities of at least 150km~s$^{-1}$ through the large scale accretion flow. This outflow is goverened by the potential of the galaxy such that the bubble rises towards the minor axis of the galaxy. The interaction between these rising bubbles of hot plasma and the large-scale accretion flow is the fundamental cause of the regions of large velocity dispersion seen in the WiFeS data.

\subsection{Accretion Shocks}
In our model, the hot plasma acts as a source of viscosity in the accreting gas, allowing the infalling gas to settle at sub-Keplerian velocity. In these circumstances, an accretion shock will form on the two faces of the dense central accretion disk-like structure. The strongly shock-enhanced emission in the accretion shock region produces the central bright disk so evident in Figure \ref{fig:NGC1052_jet}, and mapped in detail in the sub-~arcsecond resolution integral field observations of \citet{Sugai05}. These authors find that this region is characterised by a relatively slow rotation with a velocity amplitude $\sim 100-120$km~s$^{-1}$ at a radial distance of $\sim 150$pc. The inner outflow axis defined by the radio axis is strongly inclined to the rotation axis of this accretion disk by about 50$\degr$  

Shocks may also arise internally in the central accretion disk if the orbits of the accreting material are somewhat chaotic. Such a model was suggested for the case of M87 by \citet{Dopita97}. In such a case, radiative internal shocks serve to increase the binding energy of the accepted material. In other words, these shocks act as a source of viscosity in the accreted material, helping to convey gas towards the central engine. 

\subsection{Cocoon Shocks}
The mis-alignment of the radio jet axis and the accretion disk rotation axis has important consequences resulting from the interaction of the radio plasma with the (relatively thick) accretion disk. In the \citet{Sugai05} data, this interaction shows as two [\ion{O}{3}] $\lambda5007$ emission ridges associated with the \citet{Kadler04a} radio jets. The approaching (ENE) jet produces a ridge of  [\ion{O}{3}] emission blue-shifted with respect to the systemic velocity by -400~km~s$^{-1}$ out to a radial distance of about 75pc, while the receding radio jet gives a red-shifted  [\ion{O}{3}] emission ridge with a radial velocity of 200km~s$^{-1}$ out to a radial distance of $\sim100$pc. Beyond this radius, the mean radial velocity of both ridges falls rapidly towards the systemic velocity. At the same time, the  [\ion{O}{3}]/H$\beta$ ratio rises rapidly.

Here we interpret these observations in terms of the cocoon shock model for the interaction of a radio jet with an ambient interstellar medium \citep{Bicknell97,Bicknell98}. In this model, the radio jet passes though a termination shock, and becomes surrounded by a cocoon of hot plasma which is a mixture of relativistic plasma and thermal plasma entrained or swept up by the jet material. It is this cocoon which gives rise to diffuse radio emission and possibly some of the diffuse X-rays. The optical emission arises not from this component, but from a strong radiative wall shock propagating into the surrounding medium as a result of the overpressure of the cocoon material. These radiative shocks may also give rise to thermal X-ray emission if the shock velocity is high enough ($\gtrsim 200$~km~s$^{-1}$). 

In the case of NGC~1052, the mis-alignment of the jet and accretion disk axes ensures that these radiative wall shocks sit in the already shocked material of the accretion disk and are much brighter on the side of the jet closest to the mid plane of the accretion disk. From the observed radial velocities from  \citet{Sugai05}, we infer that the cocoon shock velocity is of order $\sim 300$~km~s$^{-1}$. If the direction of jet propagation is almost in the plane of the sky, then the way in which the radial velocity falls to the systemic velocity at the head of the radio jets can be explained in terms of projection. On the other hand, the increase of the [\ion{O}{3}]/H$\beta$ ratio can be ascribed to lower density, faster, and/or partially-radiative shocks at the head of the cocoon.

\begin{figure*}[htb!]
\begin{centering}
\includegraphics[scale=0.7]{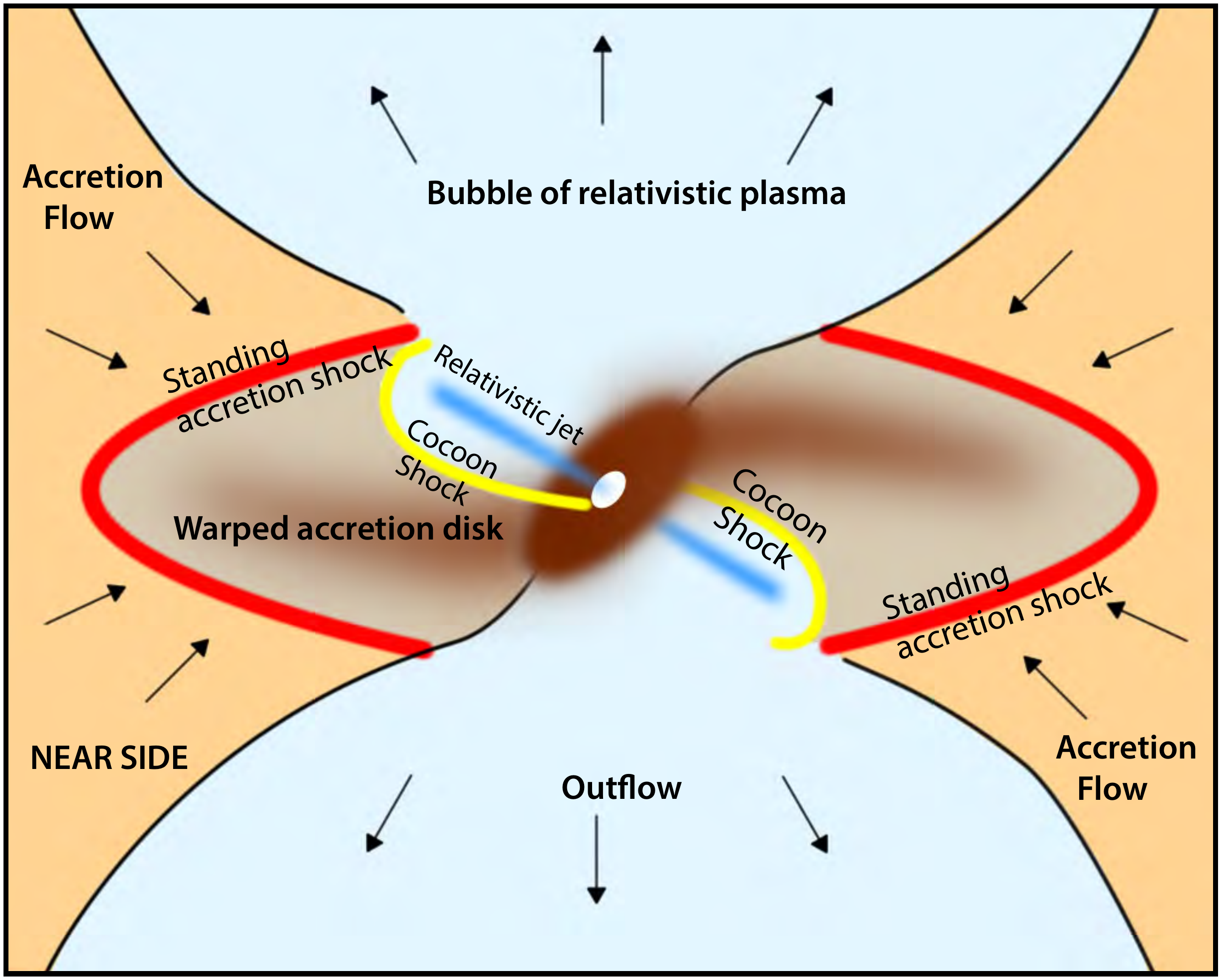}
\end{centering}
\caption{A schematic picture of the accretion flow, accretion shock and interacting cocoon shock model for NGC~1052 presented in this paper (figure courtesy of Dr. David Nicholls).}\label{fig:model}
\end{figure*}

The geometry of this model is schematically illustrated for clarity in Figure \ref{fig:model}, in which we choose a viewing angle close to the mid-plane of the galaxy. The mis-alignment of the inner accretion disk and the resulting one-sided nature of the cocoon shock is evident, as is the escape of the relativistic plasma into the expanding buoyant bubbles of gas surrounded by weak optical emission lines seen in the WiFeS observations, Figure \ref{fig:WiFeSdynamics}.

In our cocoon shock model, the radiative wall shock is propagating into a medium which has already been shocked to high density by the accretion flow. Thus, the initial density is orders of magnitude greater than the initial density in the accretion shocks. This two-shock model then would provide a natural explanation for the high- and low-density components observed to be present in the FOS observations.

We now proceed to develop this model in more detail. For this purpose we use the photoionisation and shock modelling code MAPPINGS v4.2 (Sutherland et al., in preparation). This code now incorporates a global revision of the input atomic data for all elements up to and including Zinc, as well as a full revision of the non-equilibrium ionisation/cooling solver. As such, it is more comprehensive from an atomic physics point of view, as well as being faster and more stable than previous versions of the code.

\section{Modelling the spectrum of NGC~1052}\label{sec:model}

\subsection{The Accretion Shock}
All our modelling is done with a dusty plasma having a $3\times$ solar abundance set and depletion factors as defined in \citet{Dopita13a}. Such high abundances are appropriate for the centres of massive and evolved galaxies in the local universe \citep{Zahid13}. We set up a medium in which the component of magnetic pressure support and the gas pressure are equal (equipartition). The component of the magnetic field transverse to the direction of propagation of the shocks limits the maximum compression of the post-shock gas that can be achieved. Defining the Alfv\'en Mach Number $\mathcal{M}_{\rm A}$ as the ratio of the shock velocity to the Alfv\'en velocity $v_{\rm A} =B^2/4\pi\rho_0$ where $B$ is the transverse component of the magnetic field and $\rho_0$ is the pre-shock density, the maximum compression factor in the shock,  $\rho_1/\rho_0$, is given by
\begin{equation}
\frac{\rho_1}{\rho_2} = 2^{1/2}\mathcal{M}_{\rm A}.
\end{equation}
An accretion shock produced in a medium which is initially in equipartition is already magnetically-dominated in terms of its internal pressure support by the time it has cooled enough to emit in the [\ion{S}{2}] $\lambda \lambda 6717,31$ lines. Thus, the ISM pressure  which provides the best fit to the density-sensitive [\ion{S}{2}] $\lambda \lambda 6717/6731$ ratio provides a good estimate of the initial pressure in the infalling gas for a given accretion shock velocity. On the basis of the observed velocity dispersion, we estimate the accretion shock velocity to be about 150~km~s$^{-1}$, but in order to attempt to constrain it further we investigated the effect of changing the accretion shock velocity from 80 to 260~km~s$^{-1}$.  With an accretion shock velocity of 150~km~s$^{-1}$, the parameters of the pre-shock plasma are found to be $T=10^4$K, $n_{\rm H} = 24$~cm$^{-3}$, $B=36\mu$G, $\log P/k =6.2$. Such a pressure is  not unreasonable for an Elliptical galaxy. For example, in a detailed study of five nearby relaxed giant Ellipticals \citet{Werner12} finds central pressures (in the hot X-ray plasma) averaging $\log P/k =6.5$.

\subsection{The EUV diffuse field}
For the atomic phase, we assume this to be pre-ionised by the diffuse X-ray spectrum (the molecular gas is too dense for this to be a significant source of pre-ionisation in this phase). \citet{Kadler04b} gives the absorbed luminosity of $L=(1.4)\times 10^{41}$~erg~s$^{-1}$ and an (intrinsic) luminosity in the X-rays of $L=(1.7-2.0)\times 10^{41}$~erg~s$^{-1}$. In addition, they find a very flat power-law photon index $\Gamma = 0.2 - 0.3$. The measured hydrogen absorption column density of $N_{\rm H} \sim (6-7)\times 10^{21}$~cm$^{-2}$ is consistent with a nuclear extinction of at least $A_{\rm V} \sim 3$mag. For our input radiation field we have adopted a theoretical spectrum for a black hole of $M_{\rm BH} = 10^9$M$_{\odot}$ accreting at an Eddington factor of $10^{-3}$ taken from the library of \citet{Done12,Jin12a,Jin12b,Jin12c}. The form of this is rather similar to that observed by \citet{Kadler04b}. 

\begin{figure}[htb!]
\begin{centering}
\includegraphics[scale=0.6]{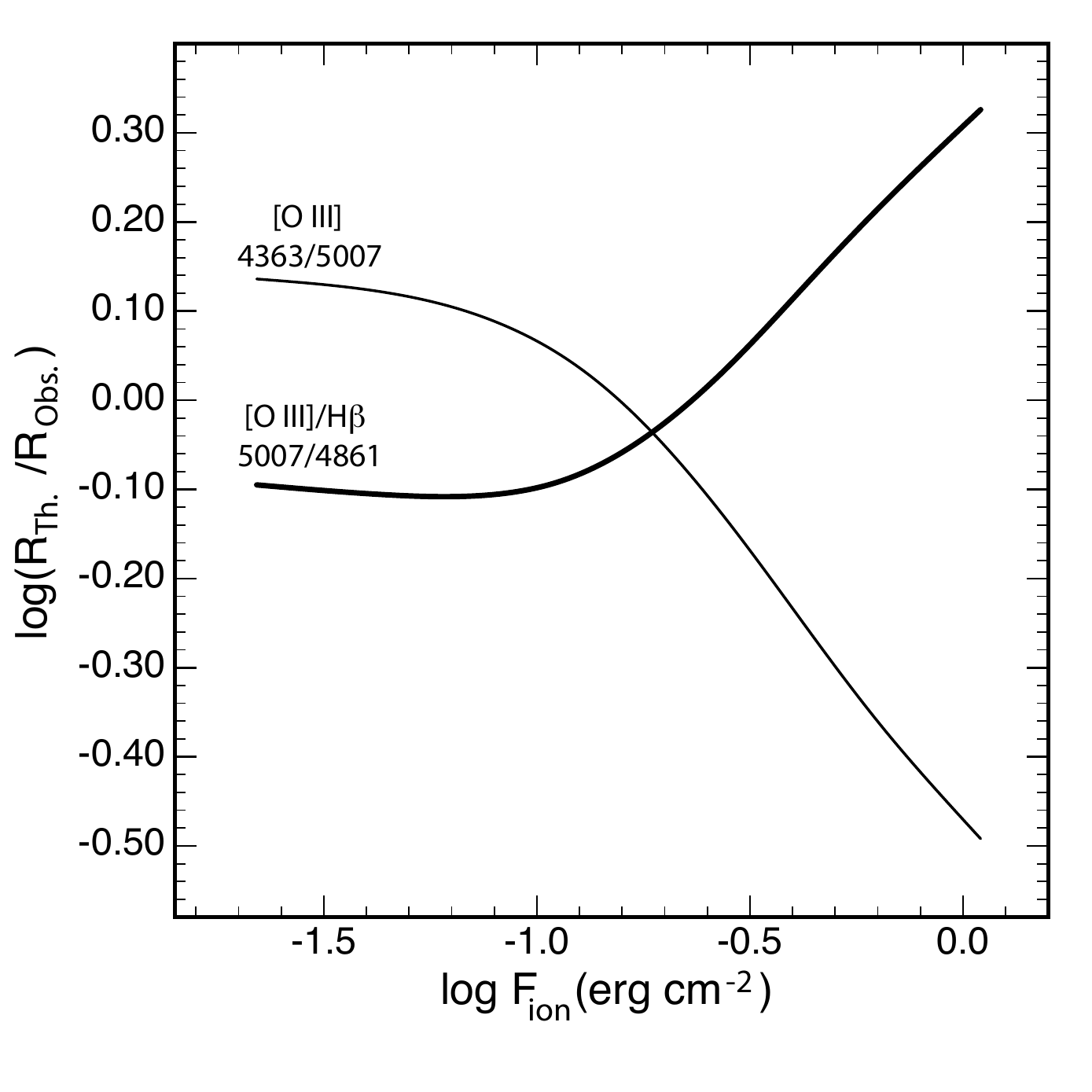}
\end{centering}
\caption{The variation of  the ratio of the theoretical to observed [\ion{O}{3}] /H$\beta$ and [\ion{O}{3}] $\lambda \lambda 4363/5007$ ratios as a function of the ionising flux in the EUV radiation field. The curves are shown for a shock velocity of $v_{\rm s} = 170$~km~s$^{-1}$ and a flux contribution from the molecular component of 0.36. These curves illustrate how tight limits can be placed on the intensity of the radiation field using the observed [\ion{O}{3}] excitation.}\label{fig:OIII}
\end{figure}

A constraint on the radiation field intensity is provided by the [\ion{O}{3}] lines. As the energy density in the radiation field comes to dominate over the energy density in the shocks, and the ionisation parameter exceeds about $\log U = -3.0$, the  [\ion{O}{3}] /H$\beta$ ratio increases to eventually approach the values typical of Seyfert galaxies. At the same time, as the ionisation parameter increases, the temperature in the  [\ion{O}{3}] zone decreases so that the  [\ion{O}{3}] $\lambda \lambda 4363/5007$ ratio decreases as the shock component of the spectrum decreases in relative importance. This is illustrated in Figure \ref{fig:OIII} for the case $v_{\rm s} = 170$~km~s$^{-1}$. In this figure, we have used the mixing fraction of the cocoon shock defined by the observations   (see Section \ref{MixingFraction}, below) The best fit to observation is obtained with  $F_{\rm X} = 0.22$~erg~cm$^{-2}$.\newline

Since the ionisation parameter characterising the X-ray photon field is determined by the geometry of the diffuse medium with respect to the AGN - which we do cannot exactly determine \emph{a priori}, we treat the dilution factor of the X-ray field as a (somewhat) free parameter. In the full family of models we have used three values of the X-ray field corresponding to $F_{\rm X} = 0.06, 0.22$ and 0.66 erg~cm$^{-2}$, which is achieved at radial distances of $2.8\times 10^{20}, 2.2\times10^{20}$ and $1.3\times 10^{20}$ cm from the central engine, respectively. This should be compared with the size of the central bright emission region in the H$\alpha$ emission, which is roughly $3\times10^{20}$~cm in radius at the distance of NGC~1052 \citep{Sugai05}.

\subsection{The Cocoon Shock}
The observations constrain the velocity of the cocoon shock to be between 200 and 300 ~km~s$^{-1}$. Since the cocoon shock is propagating into the accretion disk material which has already been processed through the accretion shock, we set the pre-shock pressure equal to post-shock pressure in the accretion shock, and once again assume equipartition. We also set the ionisation state in equilibrium with the EUV radiation field. Due to the low ionisation parameter of the local radiation field, the pre-shock gas is only partly ionised; H$^+$/H = 0.7. This gives a set of pre-shock parameters: $T_e=3000$K, $n_H=28000$~cm$^{-3}$, $B=600\mu$G, and $\log P/k =8.2$~cm$^{-3}$K. This very high pressure would have to be comparable to the pressure in the radio plasma which is driving the cocoon shock. As a result of the high pre-shock density, the electron density in the cooling zones of the shock exceeds $10^6$~cm$^{-3}$, and consequently both the cooling time and cooling length in the shock are very short, of the order of 2 years and $3\times10^{14}$cm, respectively.	

\begin{figure*}[htb!]
\centerline{\includegraphics[scale=0.8]{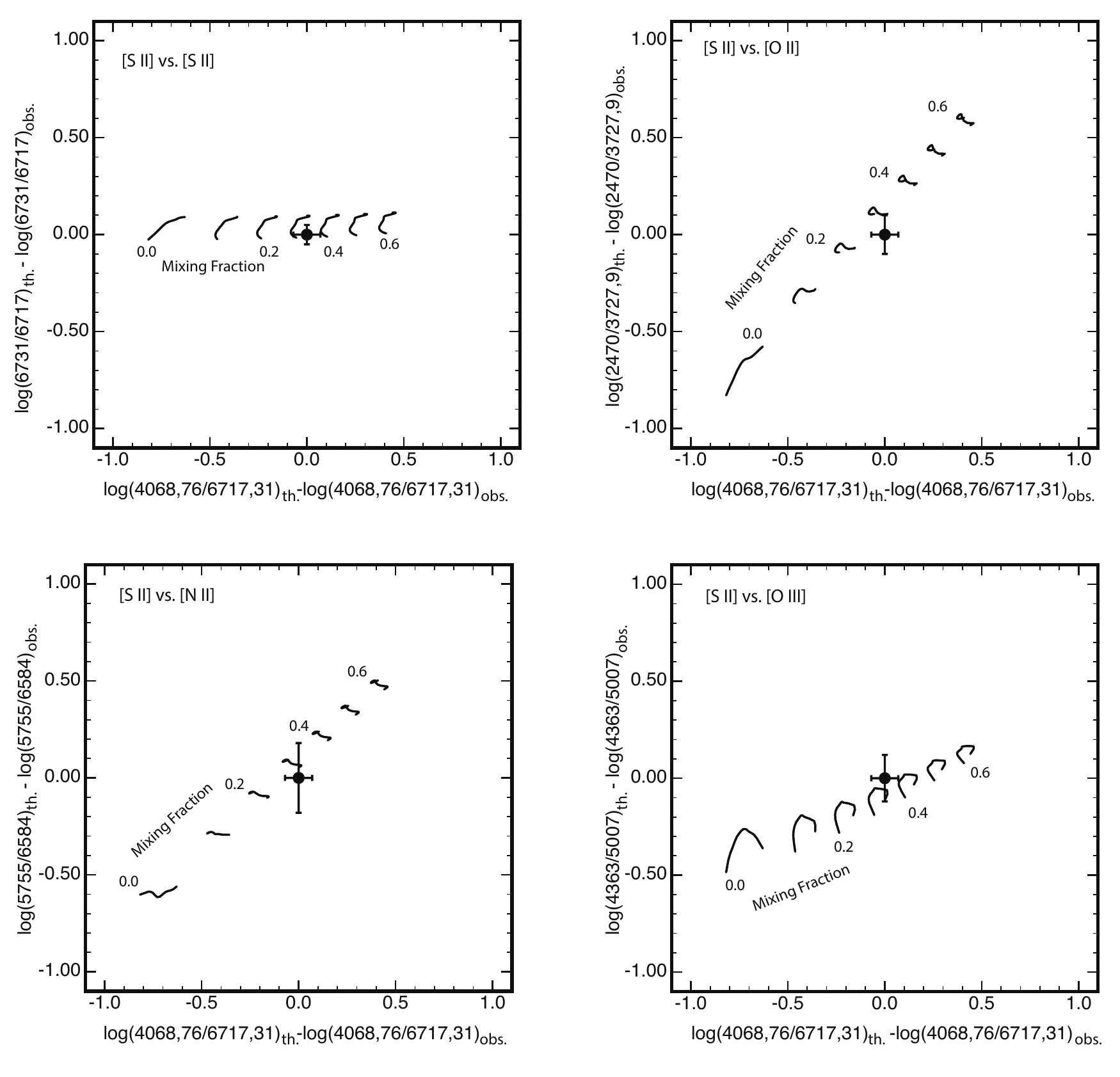}}
\caption{The variation of the density-sensitive ratios of  [\ion{S}{2}],  [\ion{O}{2}],  [\ion{N}{2}] and  [\ion{O}{3}] as a function of accretion shock velocity and mixing fraction of the cocoon shock component for a cocoon shock velocity of 200~km~s$^{-1}$. We show the ratio of the theoretical line ratio to that observed, so by definition, all the observational points are at the origin of the coordinates. The errors for the observational points are assumed to given by the photometric measurement errors. The accretion shock velocity is poorly constrained in these figures, since it varies over the range 80-200~km~s$^{-1}$ for each curve of fixed mixing fraction, but the mixing fraction itself is closely constrained by these data in the range $0.32\pm0.04$. }\label{fig:density}
\end{figure*}

\begin{figure}[htb!]
\begin{centering}
\includegraphics[scale=0.6]{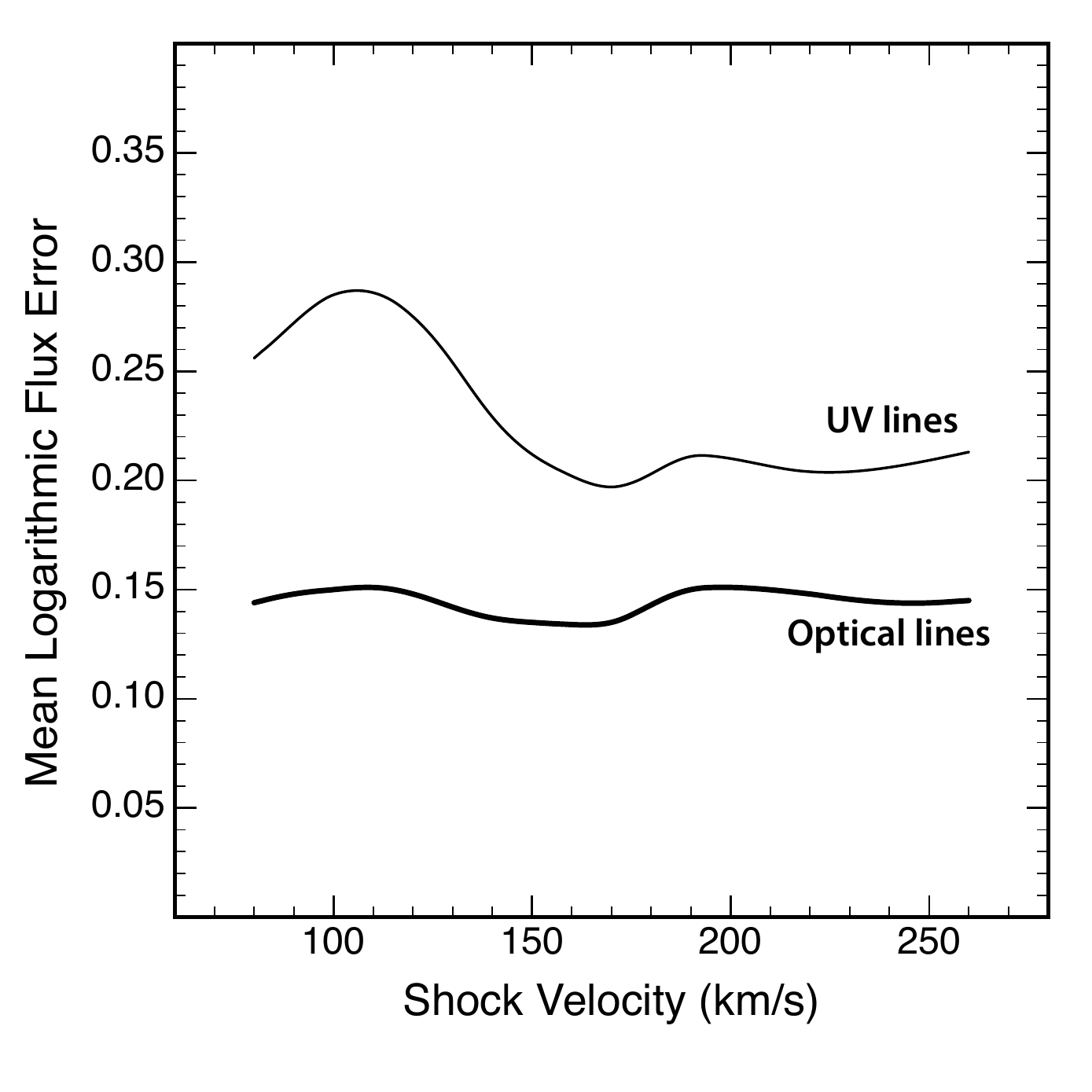}
\end{centering}
\caption{The mean absolute logarithmic difference between the line fluxes predicted by theory and the observations of NGC~1052. These are here plotted as a function of accretion shock velocity for a fixed cocoon shock velocity of 300~km~s$^{-1}$. For the UV we have fitted simultaneously to 14 lines, while for the optical the fit is for 22 emission lines. The accretion shock velocity is only weakly constrained, but for shock velocities below 130~km~s$^{-1}$ the fit is appreciably worse in the UV. This figure illustrates the insensitivity of our models to the actual accretion shock velocity. Our models are also insensitive to varying cocoon shock velocity.}\label{fig:Vel}
\end{figure}

\subsection{Constraining the Cocoon Shock Fraction, $\phi$}\label{MixingFraction}
In our model, lines such as  [\ion{S}{2}] $\lambda \lambda 6717, 6731$,  [\ion{O}{2}] $\lambda \lambda 3727, 3729$,  [\ion{N}{2}] $\lambda \lambda 6548, 6584$ and  to a lesser extent [\ion{O}{3}] $\lambda \lambda 4959, 5007$ have their origin predominantly in the accretion shock. On the other hand, lines produced in dense gas such as  [\ion{S}{2}] $\lambda \lambda 4069, 4076$,  [\ion{O}{2}] $\lambda \lambda 2470$,  [\ion{N}{2}] $\lambda \lambda 5755$ and  [\ion{O}{3}] $\lambda \lambda 4363$ as well as other lines with a high critical density such as [\ion{O}{1}] $\lambda \lambda 6300, 6363$ are produced in the cocoon shock component. This difference provides the means of determining the relative emission fraction, $\phi$, of the cocoon shocks in the combined spectrum as seen in the FOS aperture. The total flux is defined in terms of the separate fluxes from the cocoon shock and the accretion shock as $F_{\rm tot.} = \phi F_{\rm cocoon} + (1-\phi)F_{\rm accretion}$.

This difference provides a very direct way of constraining  $\phi$. In Figure \ref{fig:density} we show the sensitivity of five ratios to the shock velocity in the accretion flow and mixing fraction of the cocoon shock component. The mixing fraction is very tightly constrained, and found to lie between $0.28 \leqslant \phi \leqslant 0.35$, regardless of the shock velocity in the accretion shock which is here varied from 80 up to 200~km~s$^{-1}$. We cannot directtly estimate the mixing fraction from the observations of \citet{Sugai05}, but these are at least consistent with the fact that the cocoon shock does not dominate within the FOS aperture.

\subsection{Best Fit Models}\label{BestFit}
The best fit model is derived from a global fit to the emission line spectrum. For this purpose we compare the modulus of the logarithmic difference between theory and the observations of NGC~1052 for the predicted line intensities (relative to H$\beta$), separately for the UV and the optical lines, and attempt to minimise this difference by varying the shock velocity while keeping the molecular fraction, UV radiation field and pre-shock gas pressure constant. This procedure is known mathematically as minimising the L1 norm in the logarithmic space \footnote{see: {\tt{\tiny http://rorasa.wordpress.com/2012/05/13/l0-norm-l1-norm-l2-norm-l-infinity-norm/}}}, and it allows us to fit both bright lines and faint lines simultaneously to similar goodness of fit. 

The  result of a run with fixed cocoon shock velocity (300~km~s$^{-1}$) and varying accretion shock velocity is shown in Figure \ref{fig:Vel}. Marginally we obtain the best-fit model for an accretion shock velocity  of 150~km~s$^{-1}$. However, what is most striking about this figure is the insensitivity of the solution to accretion shock velocity -- essentially no value in the computed range is definitely excluded.

The result is also insensitive to the assumed cocoon shock velocity. All that we require is that the ram pressure of the shock is sufficient to compress the gas up to the range implied by the observed line ratios which are sensitive to high density. For illustration of this point, we include in Table 1 four models which provide an almost equally good fits to the observations. Model A and Model B both have an EUV field $F_{\rm X} = 0.07$~erg~cm$^{-2}$ and an accretion shock velocity of 150~km~s$^{-1}$, but differ in their cocoon shock velocities (200 and 300~km~s$^{-1}$, respectively). Model C and D have a stronger EUV field; 0.22~erg~cm$^{-2}$, but have the same accretion velocity (150~km~s$^{-1}$). They likewise differ in their cocoon shock velocities (200 and 300~km~s$^{-1}$, respectively). 

Note that all of these models provide a good description to the full range of ionisation states observed, ranging from \ion{O}{1} (IP = 13.61eV) up to \ion{Ne}{5} (IP = 126.2eV). A very good fit is obtained in the optical, with the exception of the [\ion{N}{1}] line - which we find to be an exceptionally difficult line to model correctly as it arises in a thin zone where hydrogen transitions from the ionised to neutral state. It is worth remarking that the intensity of the \ion{He}{2} $\lambda 4686$ line predicted by the models and given in Table 1 also looks to be in good agreement with the actual measurement of this line by \citet{Sugai05} (their Figure 1).

Comparing our results with the \citet{Gabel00} 2-phase dusty photoionisation model, we find that our model gives an appreciably improved  fit to the UV lines. This is because the high ionisation zones of our models are appreciably hotter than in photoionisation models, thanks to the influence of shock heating. In general the UV lines are enhanced in our model. Likewise the [\ion{Ne}{5}] lines in the near UV are much stronger in our model. In the red, \citet{Gabel00} note that their models underestimate the  [\ion{O}{1}] $\lambda \lambda 6300,6363$,  [\ion{N}{2}] $\lambda \lambda 6548,6584$ and  [\ion{S}{2}] $\lambda \lambda 6717,6731$ lines. Doubtless the fit would have been improved had they adopted the higher chemical abundances of these elements such as we have used in our model. Clearly a reasonable fit to the observed spectrum can be had with a 2-component photoionisation model but it would still need to be made consistent with the observed spatial and dynamical structure of NGC~1052. 

\section{Discussion}\label{discussion}
\subsection{Black Hole Mass}
The model developed in the previous section and the observations of \citet{Sugai05} can provide interesting insights on both the Black hole mass and the power in the radio jet of NGC~1052. Let us first consider what constraint can be placed on the mass of the black hole in NGC~1052 from the dynamical data.  \citet{Sugai05} quote the amplitude of the rotation velocity as 100--130~km~s$^{-1}$ at a radius of 145 pc (1.5~arcsec). The inclination of the accretion disk at this radius can only be crudely estimated as $\sim30-50$$\degr$ from the shape of the line-free continuum distribution shown in their Figure 2, or the HST H$\alpha$ observation of Figure \ref{fig:NGC1052_jet}. Assuming that the gravitational potential this close to the nucleus is dominated by the black hole, we (very crudely) estimate  $M_{\rm BH} = (4 - 14)\times10^8$M$_{\odot}$. 

An alternative approach is offered by the measurement of the central velocity dispersion in the WiFeS data of Figure \ref{fig:WiFeSdynamics}; 230~km~s$^{-1}$. Using the black hole mass: velocity dispersion relationship initially derived by \citet{Tremaine02} and refined by \citet{Graham13} would then imply a black hole mass in the range  $M_{\rm BH} = (2 -7)\times10^8$M$_{\odot}$. This is probably be the more reliable estimate of the two, although both are very crude. Nonetheless, they at least give an idea of the size of the central engine.

\subsection{Power of the Radio Jet}\label{jet}
The power of the radio jet can be estimated from the jet - cocoon theory developed by  \citet{Bicknell97}. In this, the evolution is driven by the jet parameters, the jet energy flux $\dot {E_j}$, and the relativistic $\beta=v/c$. Define the $r-$axis to be in the direction along the jets, and the $z-$axis to be the perpendicular direction. If the mean pressure in the cocoon is $P$, then towards the head of the cocoon (in the $r$ direction) the ``head'' pressure in the direction of propagation is higher than the mean pressure in the cocoon. This is caused by the relatively small area, $A$, over which the jet termination shock is jittering and the stagnation of the cocoon flow in the direction of propagation. Here we can take the pressure as a factor $\zeta$ times the average lobe pressure where, typically  $\zeta \sim$ 2 - 10. However, given that the jet appears to be more directly impinging on the cocoon of NGC~1052, we will assume  $\zeta \sim$1 for the purpose of calculation. The velocity of advance of the jet, and the velocity of the wall shocks surrounding the cavity excavated by the jet are given by, respectively:
\begin{equation}
\frac{dr}{dt} =\left[\frac{\beta\dot {E_j}}{\rho cA}\right]^{1/2} \sim \zeta^{1/2}\left[\frac{P}{\rho}\right]^{1/2} \label{eqn2}
\end{equation}
and 
\begin{equation}
\frac{dz}{dt} =\left[\frac{P}{\rho}\right]^{1/2}  \label{eqn3}
\end{equation}
where ${\rho}$ is the density in the surrounding pre-shock medium,  and $\dot {E_j}$, $\beta$, and $A$ have been defined above. The 300~km~s$^{-1}$ cocoon shock model, the post-shock pressure, $P$ is found to be $5.2\times10^{-5}$~dyne~cm$^{-2}$, and the pre-shock density, $\rho = 7.8\times10^{-20}$~gm~cm$^{-3}$.  The relativistic $\beta=0.38$ has been measured from the proper motion of hot-spots in the jets at at a radio wavelengths in the MOJAVE survey of  \citet{Lister13}. From these observations, evidence for precession in the jet is inferred. All that remains is to estimate $A$. Noting that the jet is highly collimated we assume a jet opening angle of $\lesssim 5$~degrees, which implies an area $A \lesssim 10^{38}$~cm$^2$. Substituting these into equation \ref{eqn2} gives us $\dot {E_j} \lesssim 4\times10^{44}$ erg~s$^{-1}$. This jet energy flux is comparable to those inferred for the GPS and CSS radio sources by \citet{Bicknell97}. 

\subsection{Mass Accretion Rate} 
Let us now test whether the accretion shock arises solely at an accretion shock into the outer accretion disk or has in addition contributions from internal dissipative shocks in the accretion structure itself, as suggested for the case of M87 by \citet{Dopita97}. The key parameter which the model should be capable of reproducing is the observed absolute H$\beta$ flux, which is a direct measure of the rate of increase of binding energy of the accreted gas.  From the de-reddened flux given in Table 1 and the distance of 19.5~Mpc,  we estimate $L_{\rm H \beta} \sim 4 \times 10^{39}$ erg~s$^{-1}$. The theoretical surface flux in H$\beta$ in our best fit model is $S_{\rm H\beta} =(2-4)\times10^{-3}$erg cm$^{-2}$s$^{-1}$.  The radius of the bright accretion structure from \citet{Sugai05} is about 100~pc. Therefore, the total area of the accretion disk (top and bottom) is $A \sim 5\times 10^{41}$~cm$^2$, we predict a luminosity of $L_{\rm \beta} \sim (1-2) \times 10^{39}$, in fairly good agreement with what is observed. We conclude that internal dissipative shocks within the accretion disk itself cannot contribute more than 50\% to the observed luminosity.

In the infall model the accretion rate is $\dot M = A \rho v_{\rm s}$, where $\rho$ is the mean density of the infalling gas; $\sim 5\times10^{23}$~gm~cm$^{-3}$, and $v_{\rm s} \sim 150$~km~s$^{-1}$ is the accretion shock velocity. This gives $\dot M \sim 6$~M$_{\odot}$yr$^{-1}$. This accretion rate seems remarkably high. However, we can check this by estimating the column density in the accreting material, using this to estimate the extinction it would produce, and comparing this with what is observed. For simplicity we assume that the accretion velocity remains constant, so that the accretion flow density falls off as $r^{-2}$. The hydrogen column density is therefore (approximately) the radius of the accretion disk times the pre-shock density $n_{\rm H} \sim 24$~cm$^{-3}$. This implies a column density of $\sim 7\times10^{21}$~cm$^{-2}$.  If the gas to dust ratio is similar to that of the Galaxy 
\citep{Bohlin78}, then $A_{\rm V} \sim 3$, which is a little higher than the  $A_{\rm V} \sim (1.1-1.4)$ inferred from the emission line spectrum of the central disk  (see Section \ref{FOSspectrum}) but is probably consistent with the observations when the clumpy nature of the accretion flow is taken into account. 

Although the accretion rate is high, only a small fraction of this accreted matter can find its way down to feed the central Black Hole. The radio jet parameters of NGC~1052 estimated in Section \ref{jet} imply (for a Black Hole mass of $(2 - 7)\times10^8$M$_{\odot}$) an Eddington fraction of $\lesssim (0.025 -0.1)$ and an accretion rate onto the Black Hole of $\lesssim 10^{-2}$~M$_{\odot}$yr$^{-1}$. Thus, either the radio jet is very effective in ejecting the accreted gas, or we are seeing NGC~1052 in a special epoch in which the accretion rate into the outer accretion disk is very high, while the accretion from the inner accretion disk to the Black Hole is very low.

\subsection{Comparison with other LINERs}
Although we have shown that the optical emission luminosity in NGC~1052 most likely results from accretion and mechanical luminosity in the jet.
However, we should ask whether this scenario is valid also for other LINERs, particularly those hosted by spiral galaxies, where radio luminosities and radio jets are not so common as in elliptical galaxies, and where the accretion rates could be lower.

\begin{figure}[htb!]
\centerline{\includegraphics[scale=0.5]{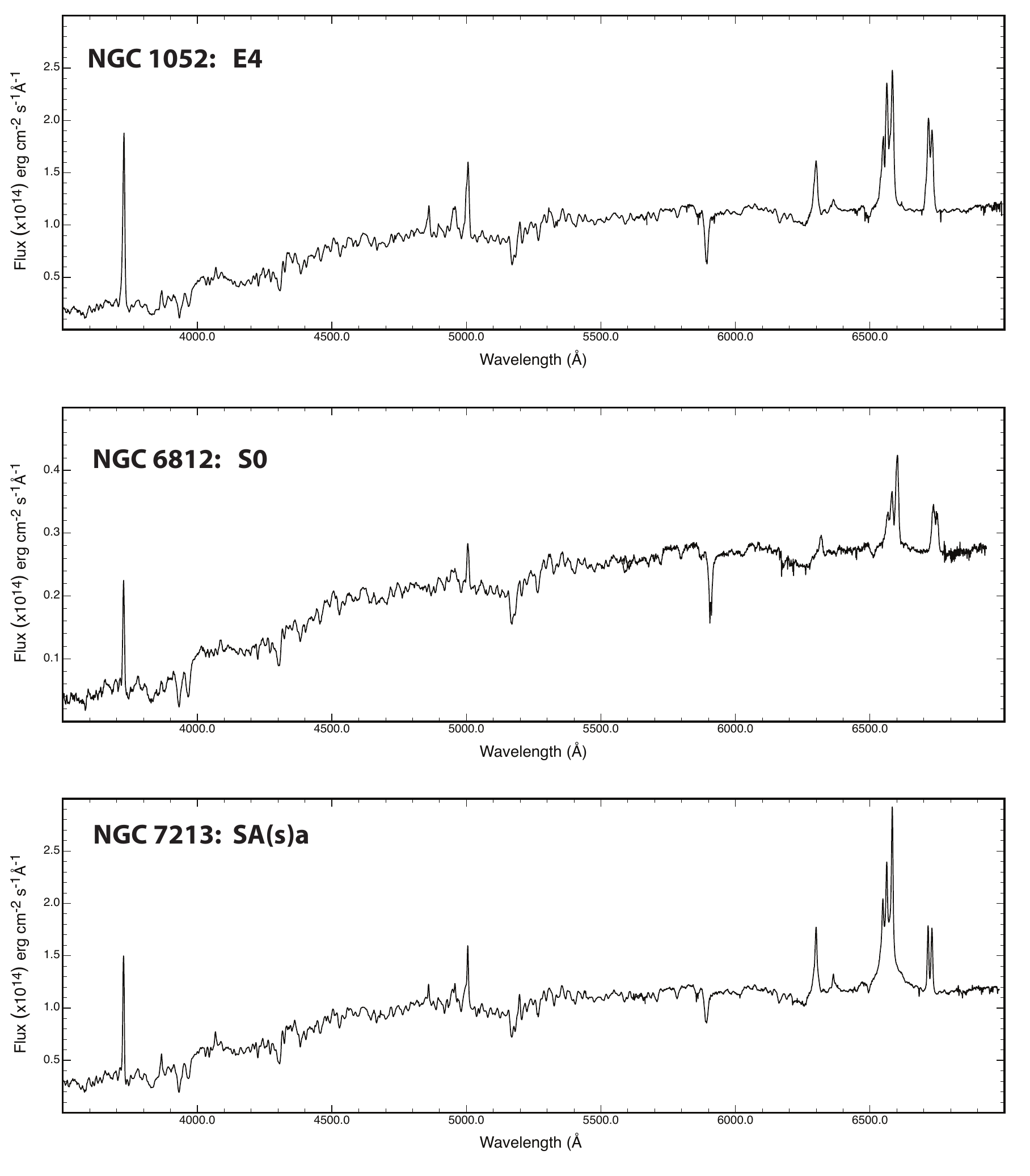}}
\caption{The WiFeS spectra of three LINER galaxies extracted from the nuclear region with an aperture of 4~arcsec. These LINERs are remarkably similar in their emission line spectrum, despite the different host galaxy types.}\label{fig:fig11}
\end{figure}

We have WiFeS data on two other galaxies, NGC~6812 and NGC~7213, taken from the S7 first data release (Dopita et al. 2015). These LINER galaxies have many features in common with NGC~1052 and cover a range of types. While NGC~1052 itself is an E4 galaxy, NGC~3031 is of SA(s)ab type, NGC~6812 is classified as an S0, and NGC~7213 is SA(s)a type in the de Vaucouleurs classification scheme. The remarkable similarity of their spectra (extracted from the nuclear region with an aperture of 4~arcsec in diameter) is shown in Figure \ref{fig:fig11}. NGC~7213 has been classified as a LINER or a Seyfert~1.5 \citep{Evans96} based upon the broad H$\alpha$ visible in  Figure \ref{fig:fig11}. However, it is clear that a broad-line LINER classification is more appropriate.

All of these galaxies - like NGC~1052 - are flat spectrum radio sources \citep{Healey07} . These exhibit significant polarisation and are believed to show appreciable Doppler boosting related to a strong jet component. In particular, NGC~7213 contains a compact and variable radio core \citep{Blank05}. This galaxy, like NGC~1052, also possesses an amorphous core emission region $\sim 130$~pc in diameter.

One of the best-studied LINERs in an spiral galaxy host is NGC~3031, better known as M81. This also contains a flat-spectrum core, with clear evidence for a precessing radio jet \citep{MartiVidal11}, which would set up ideal conditions for the type of jet-accretion disk interaction proposed here. M81 contains a black hole with a mass of about $2\times10^7$M$_{\odot}$ \citep{MartiVidal11} and references therein), which displays variable hard X-ray activity \citep{Pian10}. In the soft X-ray component there is clear evidence for thermal plasma components with two well-constrained temperatures of $0.18\pm0.04$ keV and $0.64\pm0.04$ keV \citep{Page03}. If produced by shocks, this would require shock velocities of 250 and 470~km~s$^{-1}$, respectively.

\begin{figure*}[htb!]
\begin{centering}
\includegraphics[scale=1.02]{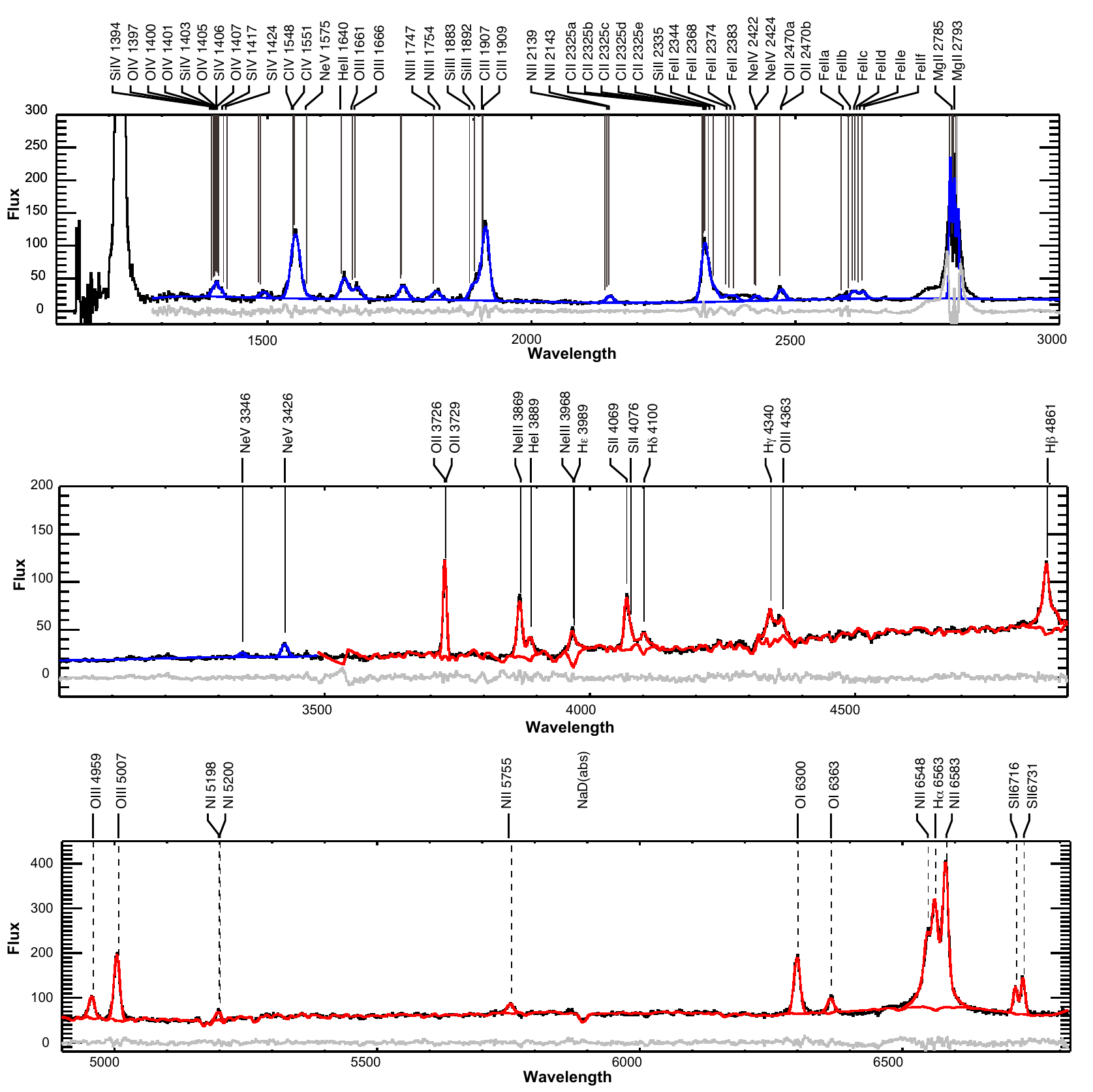}
\end{centering}
\caption{The HST FOS spectrum of NGC~3031 (solid black line), the fitted line and continuum spectrum (blue line for the UV or red line for optical wavelengths), and the residuals (gray line). Compare with Figure \ref{fig:NGC1052_HST} for NGC~1052. The spectra are very similar in all salient features, except that we see a broad line spectrum in NGC~3031 seen clearly in the \ion{Mg}{2} $\lambda 2800$ and the H$\alpha$ lines.}\label{fig:fig12}
\end{figure*}

M81, like NGC~1052, contains a compact ($\sim 140$pc diameter) disk of gas bright in \lNII\ and H$\alpha$  imaged by \citet{Devereux97} - see also \citet{Pogge00}. Leading into this disk is a spiral-like dust lane. The UV spectrum of this galaxy has been studied by \citet{Ho96} and \citet{Maoz98}. These spectra show many of the same features as NGC~1052, but with more evidence of an underlying BLR seen in \ion{Mg}{2} $\lambda 2800$ and in H$\alpha$ and H$\beta$ and in the strong featureless continuum. We also obtained high signal to noise optical-UV spectroscopic data under project GO-6532 (investigators: Dopita \& Koratkar) using  the Faint Object Spectrograph (FOS). This data is presented in Figure \ref{fig:fig12} in the same format as Figure \ref{fig:NGC1052_HST}. The similarity of the two spectra is remarkable. In particular, we see the same evidence of a two-component density structure. For example, the \ion{S}{2} $\lambda \lambda 6731/6717$ ratio in NGC~3031 is indicative of an electron density $n_e \sim 1200$, while the  \ion{S}{2} $\lambda \lambda 4069,76/6717,31$ ratio is characteristic of densities $\sim 3\times10^5$.

Regrettably, the presence of the broad-line component makes the determination of the reddening effectively impossible in this relatively low-resolution FOS spectrum, as we cannot reliably ascribe a Balmer decrement to the LINER gas. However, on the assumption that the dust extinction is negligible, we can estimate the density-sensitive line ratios, given in Table 2. These indicate densities of roughly a factor of two greater than NGC~1052 in both components of density. In the two-shock model advocated here, this is most likely accommodated by an increase in the mixing fraction of the cocoon shock (cocoon shock emission to total emission) to $0.5-0.6$ - c.f. Fig \ref{fig:density}, presumably thanks to the lower attenuation of this component. Increasing the dust extinction would drive this estimate to even higher values. The key point however is that M81 - like NGC~1052 - requires a two-density structure which is naturally explained by the two-shock model proposed here.

\section{Conclusions}
In this paper, we have investigated the complex dynamical structure of the prototypical LINER NGC~1052. We find that on the large scale (using the WiFeS instrument) ($\sim20$~arcsec), we have a pair of buoyant bubbles of plasma rising into and shocking the extended interstellar medium of the galaxies along the minor axis of the galaxy. A portion of these bubbles shows enhanced [\ion{O}{3}]/H$\beta$ line ratios, and appears to be photo ionised from the central engine, while we also see evidence for a turbulent accretion flow into a plane roughly aligned with the major axis of the galaxy, and in the sense of rotation of the stars. With HST imaging on scales of $\sim 2-4$~arcsec we see a complex dusty accretion flow and a bright central disk. This disk is resolved by the observations of \citet{Sugai05} into a rotating turbulent disk and a fast bipolar outflow which is well-aligned with the inner radio jet, but mis-aligned with the $\sim 1.0$~arcsec disk. As a consequence, the jet interacts directly with its own accretion disk.

We have built a self-consistent 2-shock model consisting of accretion shocks from an ISM with $\log P/k \sim 6.2$K~cm$^{-3}$ located at the surface of the accretion disk, and a pair of incomplete cocoon shocks which are being driven into the accretion disk by the ram pressure of the radio jet. With such a model, we can explain the observed velocity dispersion, and the full UV-optical emission line spectrum of the nuclear region of NGC~1052 obtained with FOS on the HST. The two shock model naturally explains why there seem to be two components of very different density, one with $n_e \sim10^4$ and the other with $n_e > 10^6$~cm$^{-3}$. We find that the theoretical spectrum is very sensitive to the ratio of the cocoon shock emission to the accretion shock emission, but that it is very insensitive to the assumed shock velocities, either in the accretion flow or in the jet-shocked cocoon. Thus, this type of a model may well be applicable generically to other members of the LINER family of AGN. If so, much of the optical emission luminosity of this class would be the result of accretion and mechanical luminosity in the jet, rather than due to EUV radiation from the central black hole.

\acknowledgments 
Dopita and Kewley acknowledges the support of the Australian Research Council (ARC) through Discovery project DP130103925, Dopita and Kewley acknowledges financial support from King Abdulaziz University. J.S. acknowledges the European Research Council for the Advanced Grant Program Number 267399-Momentum. SJ acknowledges support from the European Research Council through grant ERC-StG- 257720. This research has made use of NASA's Astrophysics Data System. Some of this work depends on observations made with the NASA/ESA Hubble Space Telescope, obtained from the data archive at the Space Telescope Science Institute. STScI is operated by the Association of Universities for Research in Astronomy, Inc., under NASA contract NAS 5-26555. Finally, but not least, we thank the anonymous referee for a number of constructive and valuable suggestions for improvement of the manuscript.\newpage
\bibliographystyle{apj}
\bibliography{ms}

\newpage
\begin{deluxetable}{lcccccc}
\centering
\tabletypesize{\scriptsize}
\tablecaption{Extracted nuclear line fluxes for NGC~1052 (de-reddened by A$_{\rm V} = 1.05$) compared with the models described in Section 4.5.}
\tablewidth{360pt}
\tablehead{
\colhead{$\lambda$ ({\AA})} & \colhead{Line ID}  & \colhead{Obs. Flux$^{1}$ } & \colhead{ Model A$^{2}$ } & \colhead{ Model B$^{3}$} & \colhead{ Model C$^{4}$} & \colhead{ Model D$^{5}$ }\\
\colhead{} & \colhead{} & \colhead{(H$\beta=100$)} & \colhead{} & \colhead{} & \colhead{} & \colhead{} \\
}
%\begin{table*} \label{table:FOS}
%\centering
%\caption{Extracted nuclear line fluxes for NGC~1052 (de-reddened by A$_{\rm V} = 1.05$) compared with the models described in detail in Section 4.5.}
%\begin{tabular}{lcccccc}
%\tableline \tableline
%$\lambda$ ({\AA}) & Line ID & Flux &  Model A  & Model B & Model C & Model  D\\
% &  & (H$\beta=100$) &  $200$~km~s$^{-1}$ &$200$~km~s$^{-1}$ &$300$~km~s$^{-1}$ \\
%\tableline
\startdata
1548,51$^{6}$& \ion{C}{4}     &  $72 \pm 12$ \footnote{Errors given are measurement errors only. \\Errors due to reddening correction are not computed}  & 174 & 119 & 144 & 123\\
1640      & \ion{He}{2}   & $74 \pm9$     &  72& 69 & 73 & 79 \\
1661,66 & [\ion{O}{3}]    & $30 \pm 10$  &  67 & 57 & 70 & 60 \\
1883,92 & \ion{Si}{3}]   &  $51  \pm 7$   & 12 & 10 & 13 & 12 \\
1907,09  & \ion{C}{3}]   &  $129 \pm 7$ & 69 & 67 & 82 & 92 \\
2139,43  & \ion{N}{2}]   &   $38 \pm 5$  &  8 & 8 & 8 & 8 \\
2323,35  &\ion{C}{2}], \ion{Si}{2}]  &  $107 \pm 13$ & 116 & 134 & 116 & 122 \\
2470       & [\ion{O}{2}]     & $41 \pm 3$  & 58 & 54 & 82 & 89 \\
2422,24  & [\ion{Ne}{4}]   &  $37 \pm 4$ & 30 & 27 & 28 & 24 \\
2785,93  & \ion{Mg}{2}    & $67 \pm 6$ & 55& 66 & 45 & 48 \\
3345       & [\ion{Ne}{5}]   & $4  \pm 2$  & 5 & 5 & 5 & 4 \\
3426       & [\ion{Ne}{5}]   &  $12 \pm 3$ & 13 & 12 & 15 & 12 \\
3726,29  & [\ion{O}{2}]     & $ 403 \pm 8$ & 595 & 617 & 665 & 647 \\
3869       &[\ion{Ne}{3}]    &  $76 \pm 8$ & 59 & 65 & 71 & 97 \\
3889       &\ion{He}{1},\ion{H}{1}    &   $26 \pm 2$ & 26 & 28 & 26 & 27 \\
3968,70  & [\ion{Ne}{3}], H$\epsilon$ &  $39 \pm 8$ & 34 & 36 & 35 & 43 \\
4069,76  &[\ion{S}{2}]      & $77 \pm 5$ & 109 & 121 & 84 & 80 \\
4100       &H$\delta$        & $19 \pm 3$ & 26 & 26 & 26 &  26 \\
4340       &H$\gamma$   &  $50 \pm 3$ & 46 & 47 & 47 & 47 \\
4363       & [\ion{O}{3}]     &  $21 \pm 3$ & 16 & 13 & 16 & 15 \\
4686       &\ion{He}{2}      &   ...  & 10 & 10 & 10 & 11 \\
4861       & H$\beta$        &  $100\pm 3$ & 100 & 100 & 100 & 100 \\
4959       &[\ion{O}{3}]      &  $85 \pm 3$  & 73 & 64 & 78 & 76 \\
5007       & [\ion{O}{3}]     & $241\pm 4$ & 211 & 186 & 225 & 220 \\
5198,200 &[\ion{N}{1}]    & $31 \pm 4$ & 2 & 3 & 2 & 3 \\
5755       &[\ion{N}{2}]      &  $12 \pm 5$ & 13 & 14 & 15 & 17 \\
5876       & \ion{He}{1}]    &  $17 \pm 6$ & 20 & 22 & 20 & 21 \\
6300      & [\ion{O}{1}]      &  $183\pm 5$ & 183 & 241 & 121 & 175 \\
6363      & [\ion{O}{1}]      &   $54 \pm 3$ & 58 & 77 & 39 & 56 \\
6548      & [\ion{N}{2}]      &  $111 \pm 7$ & 110 & 110 & 126 & 123 \\
6563      & H$\alpha$      & $274 \pm 20$& 294 & 294 & 294 & 294 \\
6584      & [\ion{N}{2}]      & $330 \pm 21$ & 322 & 324 & 372 & 362 \\
6716      & [\ion{S}{2}]      & $143 \pm 5$ & 190 & 188 & 184 & 186 \\
6731      & [\ion{S}{2}]      & $131 \pm 5$ & 169 & 178 & 195 & 190 \\
\enddata
\tablenotetext{1}{~Absolute de-reddened flux in aperture:  $F(H_\beta) = 9.04\times10^{-14}$ (erg cm$^{-2}$s$^{-1}$).}
\tablenotetext{2}{$F_{\rm X} = 0.07$~erg~cm$^{-2}$, accretion shock velocity 150~km~s$^{-1}$, cocoon shock velocity 200~km~s$^{-1}$.}
\tablenotetext{3}{$F_{\rm X} = 0.07$~erg~cm$^{-2}$, accretion shock velocity 150~km~s$^{-1}$, cocoon shock velocity 300~km~s$^{-1}$.}
\tablenotetext{4}{$F_{\rm X} = 0.22$~erg~cm$^{-2}$, accretion shock velocity 150~km~s$^{-1}$, cocoon shock velocity 200~km~s$^{-1}$.}
\tablenotetext{5}{$F_{\rm X} = 0.22$~erg~cm$^{-2}$, accretion shock velocity 150~km~s$^{-1}$, cocoon shock velocity 300~km~s$^{-1}$.}
\tablenotetext{6}{~Line doublets are indicated thus (i.e) 1548,51 = $\lambda1548+\lambda1551$ }
\end{deluxetable}

%\tableline
%\end{tabular}
%\end{table*}
%Parameters of models v_accn = 150km/s: A,B  Dilf=3e-13.  C,D Dilf=1e-12 (=best fit to EUV specific intensity)

\begin{table} \label{table:Ratios}
\centering
\caption{A comparison of the density-sensitive line ratios measured with FOS in  NGC~1052 and in NGC~3031.}
\begin{tabular}{lccc}
\tableline \tableline
Line Ratio & Ion & NGC~1052 &  NGC~3031\\
 \tableline
6731/6717 & [\ion{S}{2}]    & $0.91\pm0.06$  &  $1.32\pm0.06$ \\
4069,76/6717,31 & [\ion{S}{2}]    & $0.28\pm0.05$  &  $0.53\pm0.09$ \\
4363/5007 & [\ion{O}{3}]    & $0.087\pm0.012$  &  $0.149\pm0.049$ \\
2470/3726,9 & [\ion{O}{2}]    & $0.107\pm0.009$  &  $0.30\pm0.05$ \\
5755/6584 & [\ion{N}{2}]    & $0.027\pm0.013$  &  $0.107\pm0.011$ \\
\tableline
\end{tabular}
\end{table}

\end{document}